%% file: VQC_QRL_main.tex
\documentclass[a4paper,twoside]{article}

\usepackage{epsfig}
\usepackage{subcaption}
\usepackage{calc}
\usepackage{amssymb}
\usepackage{amstext}
\usepackage{amsmath}
\usepackage{amsthm}
\usepackage{multicol}
\usepackage{pslatex}
\usepackage{apalike}
\usepackage[bottom]{footmisc}
\usepackage{tikz}
\usetikzlibrary{quantikz}
\usepackage{siunitx}
\usepackage{SCITEPRESS}     

\begin{document}
\title{Quantum Reinforcement Learning for Solving a Stochastic Frozen Lake Environment and the Impact of Quantum Architecture Choices}

\author{\authorname{Theodora-Augustina Drăgan\sup{1}, Maureen Monnet\sup{2}, Christian B. Mendl\sup{3} and Jeanette M. Lorenz\sup{4}}
\affiliation{\sup{1,2,4}Fraunhofer Institute for Cognitive Systems IKS, Munich, Germany}
\affiliation{\sup{3}Technical University of Munich, Department of Informatics, Boltzmannstraße 3, 85748 Garching, Germany and Technical University of Munich, Institute for Advanced Study, Lichtenbergstraße 2a, 85748 Garching, Germany}
\email{\sup{1}theodora-augustina.dragan@iks.fraunhofer.de, \sup{2}maureen.monnet@iks.fraunhofer.de, \sup{3}christian.mendl@tum.de, \sup{4}jeanette.miriam.lorenz@iks.fraunhofer.de}
}


\abstract{Quantum reinforcement learning (QRL) models augment classical reinforcement learning schemes with quantum-enhanced kernels. Different proposals on how to construct such models empirically show a promising performance. In particular, these models might offer a reduced parameter count and shorter times to reach a solution than classical models. It is however presently unclear how these quantum-enhanced kernels as subroutines within a reinforcement learning pipeline need to be constructed to indeed result in an improved performance in comparison to classical models. In this work we exactly address this question. First, we propose a hybrid quantum-classical reinforcement learning model that solves a slippery stochastic frozen lake, an environment considerably more difficult than the deterministic frozen lake. Secondly, different quantum architectures are studied as options for this hybrid quantum-classical reinforcement learning model, all of them well-motivated by the literature. They all show very promising performances with respect to similar classical variants. We further characterize these choices by metrics that are relevant to benchmark the power of quantum circuits, such as the entanglement capability, the expressibility, and the information density of the circuits. However, we find that these typical metrics do not directly predict the performance of a QRL model.}

\onecolumn \maketitle \normalsize \setcounter{footnote}{0} \vfill


\input{chapters/01_introduction}
\input{chapters/02_background}
\input{chapters/04_task_and_solution}
\input{chapters/05_results}
\input{chapters/03_quantum_metrics}
\input{chapters/06_conclusions}
\section*{\uppercase{Acknowledgements}}
The research is part of the Munich Quantum Valley, which is supported by the Bavarian state government with funds from the Hightech Agenda Bayern Plus.
\bibliographystyle{apalike}
{\small
\bibliography{VQC_QRL_main}}
\end{document}

%% file: chapters/01_introduction.tex
\section{INTRODUCTION}
\label{sec:introduction}

Reinforcement learning (RL) is one of the pillars of machine learning, together with supervised and unsupervised learning. It has many industry-relevant applications, such as robotic tasks on assembly lines \cite{christiano2016transfer}, drug design \cite{popova2018deep}, and navigation tasks \cite{zhu2017target}. When applying RL to complex settings and environments, there is currently a tendency to focus on approximate solutions due to the complexity involved. In this context, the use of (deep) neural networks (NN) as value function approximators inside the policy of the RL agent has gained in popularity. Here, the state of the environment is processed by a NN in order to attribute a value function to it, or to estimate the value of each possible action to be taken from that state. This leads to policy gradient algorithms such as Deep Q-Networks (DQN) \cite{mnih2013playing} and Proximal Policy Optimization (PPO) \cite{schulman2017proximal}, which can even play video games such as ``Space Invaders" \cite{mnih2013playing}. The issues with these approaches, however, is that they use increasingly deep NNs, which may take several days of training on a GPU for one problem instance~\cite{ceron2021revisiting}. With increasingly complex environments, these methods may therefore experience scaling issues. Hence, it is interesting to explore alternative methods and directions. 

The use of quantum computing (QC) subroutines could be a promising path for RL. QC has theoretically been shown to exponentially or polynominally accelerate important subroutines, in particular for search problems via Grover's algorithm \cite{grover1996fast} or for solving systems of linear equations with the HHL algorithm \cite{harrow2009quantum}. Moreover, theoretical work \cite{caro2022generalization} shows that certain quantum algorithms and quantum-enhanced models may lead to a better generalization than classical algorithms in the case of a small training dataset. Further work hints that quantum algorithms may be able to observe new non-trivial characteristics in the data \cite{liu2021rigorous}, as well as to reach similar or better results, while requiring less training steps \cite{heimann2022quantum}. 

Different successful proposals have been made on how quantum algorithms could be integrated in RL models. These can be distinguished into two main directions, either trying to employ quantum search algorithms as replacement for the agent \cite{niraula2021quantum}, or replacing the NN part with a quantum circuit \cite{heimann2022quantum}. Possible application fields include robot navigation tasks~\cite{heimann2022quantum} or medical tasks. E.g., within the oncological area, a proposal was made to adapt the radiotherapy treatment plans of lung cancer patients depending on how they have responded to previous treatments~\cite{niraula2021quantum}. However, none of the existing works in quantum reinforcement learning (QRL) have investigated the correlation between the performance of the solution and the architecture of the quantum circuit yet. It is indeed unclear from the literature what more or less promising architectures for quantum kernels within a RL model might be.

Within this work, we investigate multiple additions to the field. First, we define a variant of the slippery frozen lake (FL) example~\cite{brockman2016gym}, which is a significantly more difficult stochastic environment in comparison to the typically used deterministic FL. We then consider a variety of different hybrid quantum-classical (HQC) PPO models, where the NNs of the classical algorithm are replaced by parametrised quantum circuits (PQC). We obtain good performance in comparison to the classical models considered, with respect to the number of time steps required until training convergence and the maximal reward reached during training. The quantum circuits within these models are chosen following suggestions from the literature about promising quantum circuits \cite{sim2019expressibility}, which e.g., are more efficient in using present-term quantum hardware, or may solve specific subproblems. We then characterise the solution with quantum-related metrics such as entanglement capability, expressibility, and effective dimension. We find that although all quantum circuits show a promising performance, the gain in performance does not seem to directly correlate to these quantum metrics. Therefore, our work can only give a first indication for promising quantum architectures in QRL.

This paper is structured as follows: the next section will present the related work and current status of using quantum computing in RL. The third section introduces the slippery FL environment and the pipeline of our PPO-based solution. Section four details the results achieved by our HQC PPO model. The fifth section first defines the three quantum metrics used, and then looks at the correlation with the results obtained. Finally in the sixth section we present conclusions and propose possible future research directions.

%% file: chapters/02_background.tex
\section{RELATED WORK}
\label{sec:background}

In the stream of HQC RL solutions that use quantum search algorithms, a model in the oncological area has been presented by \cite{niraula2021quantum}. The goal is to adapt the last two weeks of a radiotherapy treatment plan depending on the patient's reaction to the first four weeks of treatment. The need for artificial intelligence methods comes from the fact that numerous biological factors can influence the individual impact of the treatment. The proposed solution uses Grover's search to decide the next step in the course of the treatment by taking into account the current response. Oncological metrics were evaluated in order to maximise the treatment outcome and optimise its efficiency. In this particular case, QC is not used to accelerate the calculation, but to improve the precision.

A second direction in QRL considers replacing NNs by quantum circuits. Such methods have shown an interesting potential to solve complex tasks while requiring less computational resources than classical RL models \cite{jerbi2021parametrized,chen2020variational,lockwood2020reinforcement}. In these cases, quantum circuits are used for approximating both policy and value functions. 

In~\cite{chen2020variational}, the authors suggest using PQCs instead of NNs in the DQN algorithm~\cite{mnih2015dqn}. The expectation values of the PQC are measured and associated to the Q function of the DQN algorithm. They obtain promising results in solving a shortest-path deterministic FL and a Cognitive Radio \cite{gawlowicz2019ns} environment.

Alternatively, \cite{jerbi2021parametrized} replace the policy of a classical RL algorithm \cite{sutton2018reinforcement} with a PQC, while computing the value function classically. This work also employs the idea of a data re-uploading circuit \cite{perez2020data} in the PQC, where data embedding and variational parts are sequentially repeated to increase the overall complexity of the PQC. The authors benchmark two different hybrid quantum architectures against the maximal reward obtained during learning on Gym environments \cite{brockman2016gym}. While this work shows a promising hybrid advantage on two environments, it is not thoroughly tested against classical solutions. Nevertheless, one can clearly observe how architectural choices and hyperparameters such as circuit depth can impact the performance.

A hybrid data re-uploading technique can also help a robot agent to navigate through a maze-like environment and reach a desired solution, while avoiding obstacles \cite{heimann2022quantum}. In this work, both of the NN function approximators of the Double Q-learning algorithm \cite{van2016deep} were replaced with a hybrid module containing a PQC followed by a post-processing layer. The results of the hybrid solution are competitive with the ones a classical solution, reaching the reward threshold after a similar number of training steps. Moreover, the hybrid solution needs 10 times fewer trainable parameters to solve the environment.

%% file: chapters/04_task_and_solution.tex
\section{\MakeUppercase{Solving a slippery frozen lake environment}}
\label{sec:task_solution}
\begin{figure*}[!h]
    \centering
    \includegraphics[scale = 0.34]{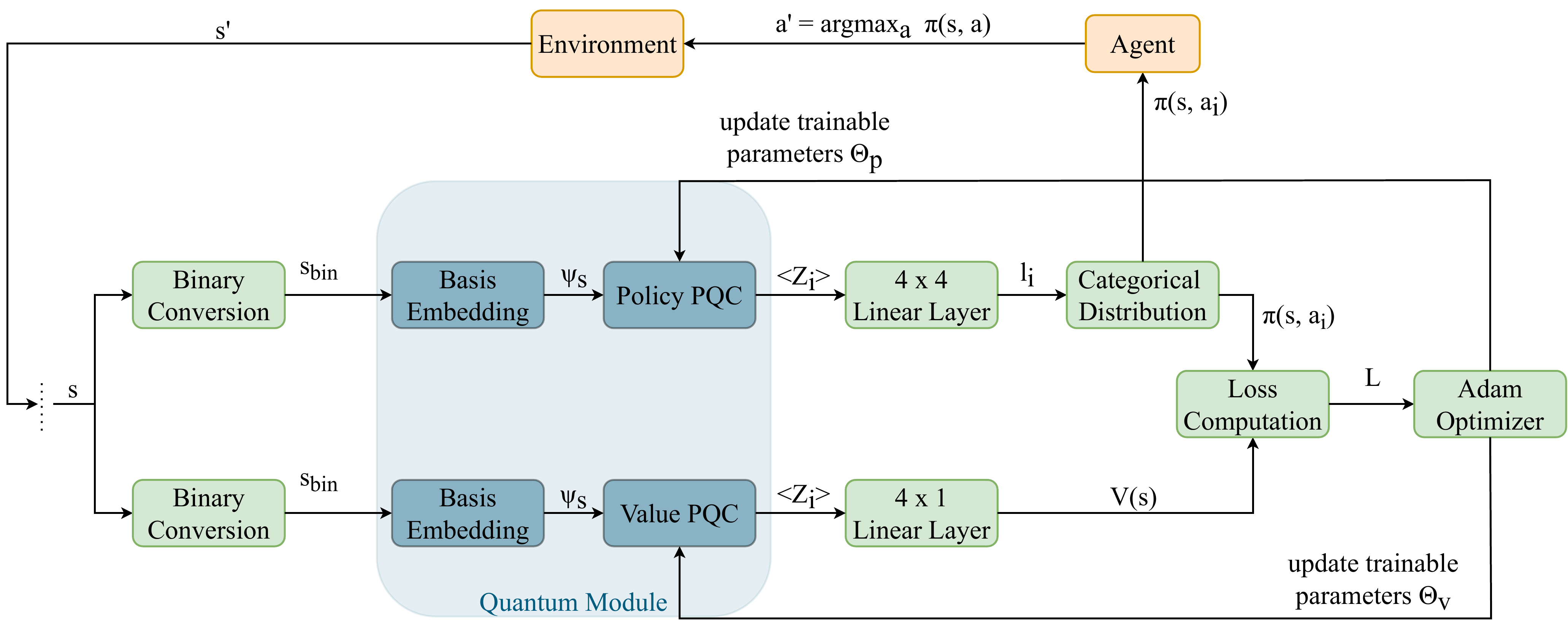}
    \caption{The entire pipeline of our solution.}
    \label{fig:pipeline}
\end{figure*}

\subsection{The Slippery Frozen Lake}
\begin{figure}[!h]
    \centering
    \includegraphics[scale = 0.35]{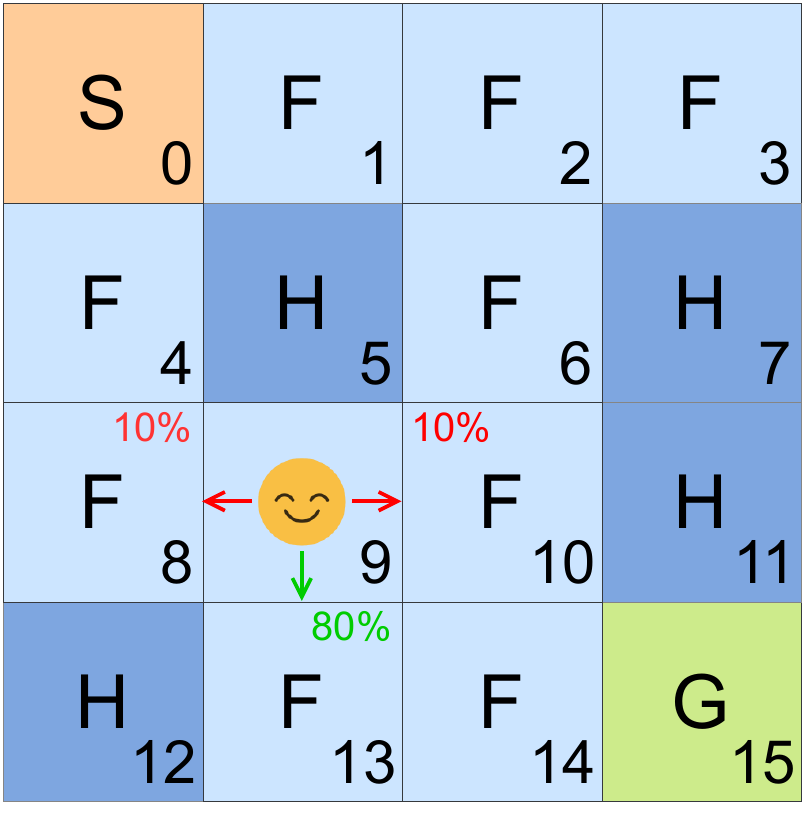}
    \caption{The structure of the FL for the state $ s = 9 $.}
    \label{fig:frozen-lake}
\end{figure}

The FL environment is a common benchmark example in classical RL \cite{steckelmacher2019sample,khadke2019exploration,gupta2021evaluation}. The task consists of navigating through a maze e.g., of size 4x4 in this study. The maze contains four types of discrete tiles, including the start (S), the goal (G), the tiles with holes (H) to be avoided, and the frozen tiles the agent can walk across (F). A previous example of solving this environment by QRL reduced the task to a deterministic environment, where the agent navigating through the maze would always move into the direction desired \cite{chen2020variational}. We found that even for present small QRL models, this is a fairly simple task to solve. Therefore, we opted for a significantly more difficult slippery FL environment~\cite{brockman2016gym}, which is stochastic in nature. In the default stochastic FL, the probability to slip and consequently to move to an orthogonal direction to the one desired is $\frac{2}{3}$. To mimic real environments more closely, we reduced the probability to slip in this study to 20\%, as shown in Fig. \ref{fig:frozen-lake}.

The tiles and thus the possible states of the environment are integers from 0 to 15. From any position, the agent can go left, down, right or up. If the agent moves to a hole, it receives a final reward of 0 and the learning process terminates. If the agent reaches the goal, it receives a reward of 1. Moving across the lake on the S and F tiles results in a 0 reward. 

In order to control the quality of the agent that learnt to navigate through an environment, usually a reward threshold is established. If the reward obtained by the agent is equal to or surpasses a given threshold, the agent is considered to have successfully learnt the environment. 
The reward threshold is calculated based on the average reward obtained when following the optimal policy. Therefore, we first determined the optimal policy for our slippery FL environment and averaged its rewards over 1000 test episodes. With the resulting average reward of 0.85, we deducted a reasonable reward threshold of 0.81, in a similar way as done in the FL documentation \cite{brockman2016gym}.

\subsection{The Architecture of the HQC Solution}

To solve our slippery FL environment, we designed a hybrid PPO algorithm. The PPO algorithm was chosen as a basis since it achieved the highest average rewards for many Gym environments~\cite{schulman2017proximal}, in comparison to other policy gradient methods. The PPO implementation used was provided by the Stable Baselines 3 Python library \cite{raffin2021stable}.

The first experiments were performed without any quantum kernel to create classical baselines. Here, the policy and value function approximators are NNs with two hidden layers of either 2, 4, 8, and 16 neurons each, followed by a post-processing linear layer. The relatively low number of neurons was chosen to enable a fairer comparison with the hybrid variants with quantum kernels, as these require a relatively small number of trainable parameters. The input to the classical solution is an one-hot vector encoding of the state. 

In the HQC solution, the algorithm remains the same except for the policy and the value NNs, which are replaced by quantum kernels, as shown in Fig. \ref{fig:pipeline}. The quantum kernels consist of a data encoding, the trainable PQC, a measurement step, and a classical post-processing. The data encoding is done via basis embedding using $R_X$ and $R_Z$ gates. The state index $s$ is transformed into its binary representation $s_{bin} = s_0s_1s_2s_3$, and is used to turn the default $\ket{0000}$ input state into the $\psi_s = \ket{q_0q_1q_2q_3}$ state. This is achieved by setting the rotational parameters of the $R_X(\theta)$ and $R_Z(\theta)$ gates acting on wire $i$ to $\theta = 0$, if $s_i = 0$, and to $\theta = \pi$ if $s_i = 1$. This $\psi_s$ state is the input to the trainable parts of the circuits, which are the Policy PQC and Value PQC of Fig. \ref{fig:pipeline}. 

\begin{figure*}[!h]
\captionsetup[subfigure]{labelformat=empty}
\centering
\subfloat[Circuit 1]{
\includegraphics[height=0.13\textwidth]{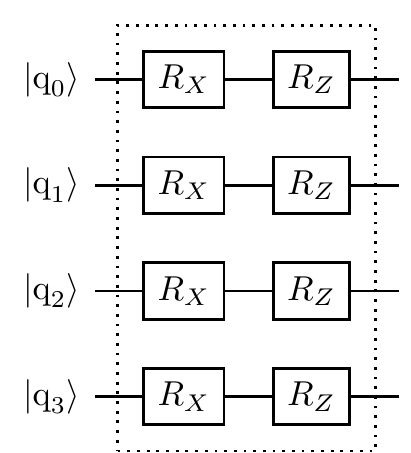}
\label{fig:circ-1}}
\subfloat[Circuit 2]{
\includegraphics[height=0.13\textwidth]{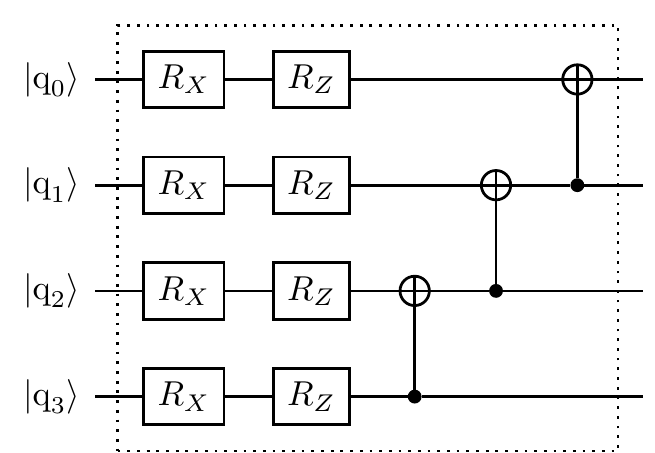}
\label{fig:circ-2}}
\subfloat[Circuit 3]{
\includegraphics[height=0.13\textwidth]{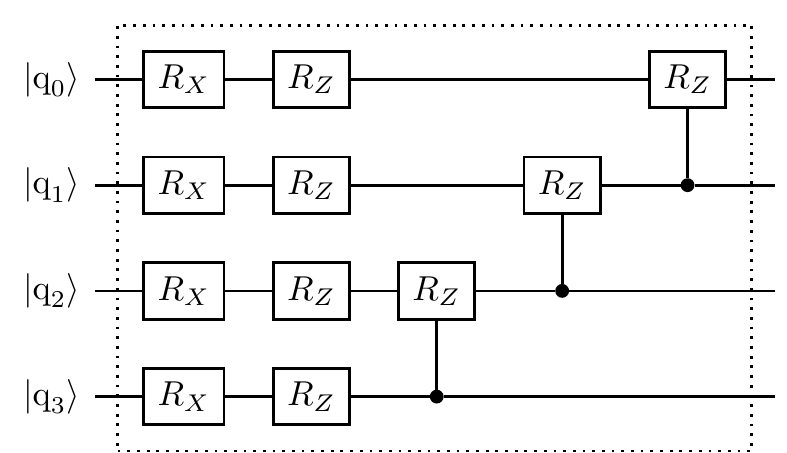}
\label{fig:circ-3}}
\subfloat[Circuit 4]{
\includegraphics[height=0.13\textwidth]{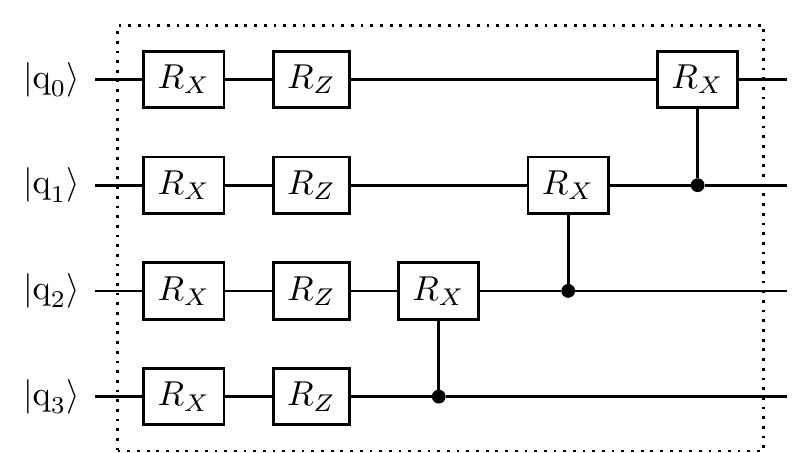}
\label{fig:circ-4}}\\~\\~\\
\subfloat[Circuit 5]{
\includegraphics[height=0.13\textwidth]{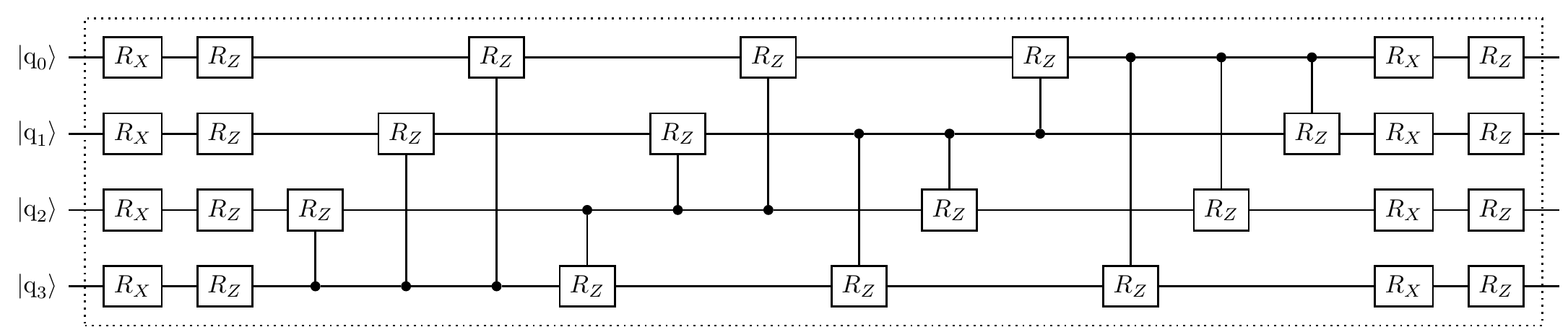}
\label{fig:circ-5}}
\subfloat[Circuit 7]{
\includegraphics[height=0.13\textwidth]{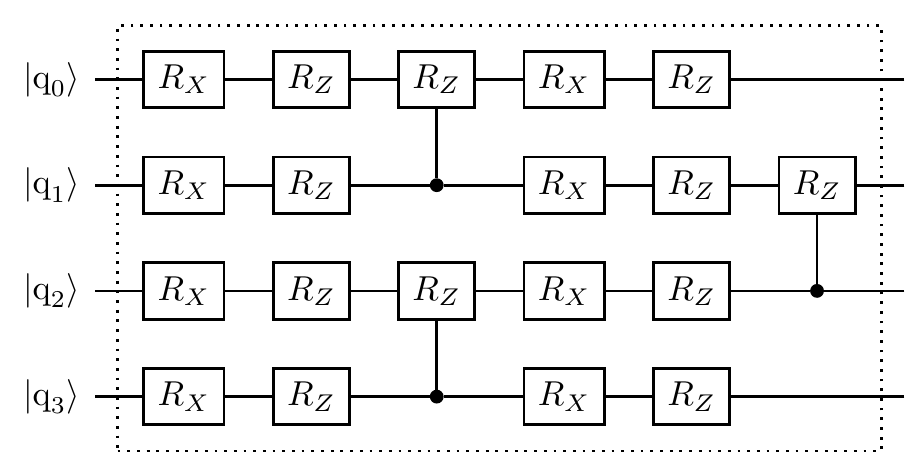}
\label{fig:circ-7}}\\~\\~\\
\subfloat[Circuit 6]{
\includegraphics[height=0.13\textwidth]{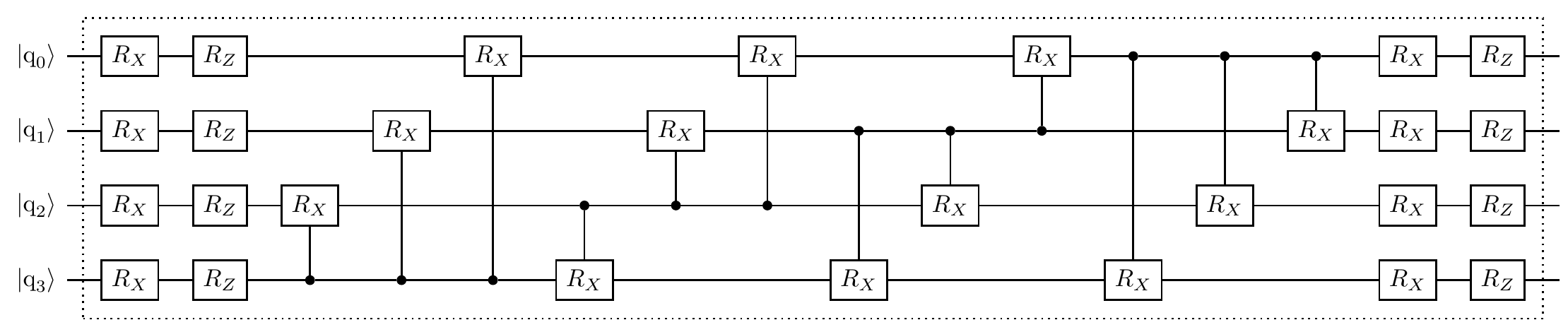}
\label{fig:circ-6}}
\subfloat[Circuit 9]{
\includegraphics[height=0.13\textwidth]{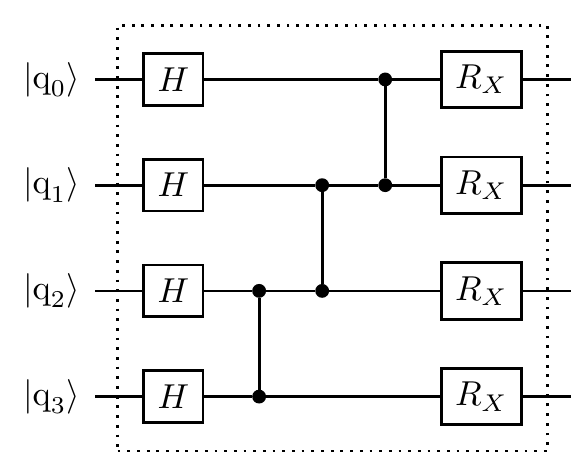}
\label{fig:circ-9}}
\subfloat[Circuit 10]{
\includegraphics[height=0.13\textwidth]{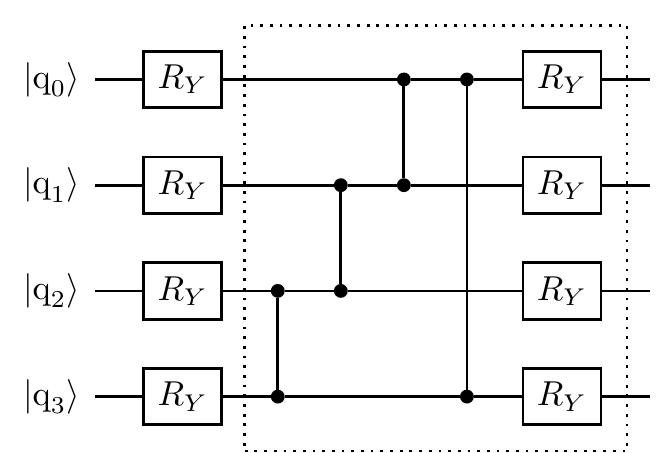}
\label{fig:circ-10}}\\~\\~\\
\subfloat[Circuit 8]{
\includegraphics[height=0.13\textwidth]{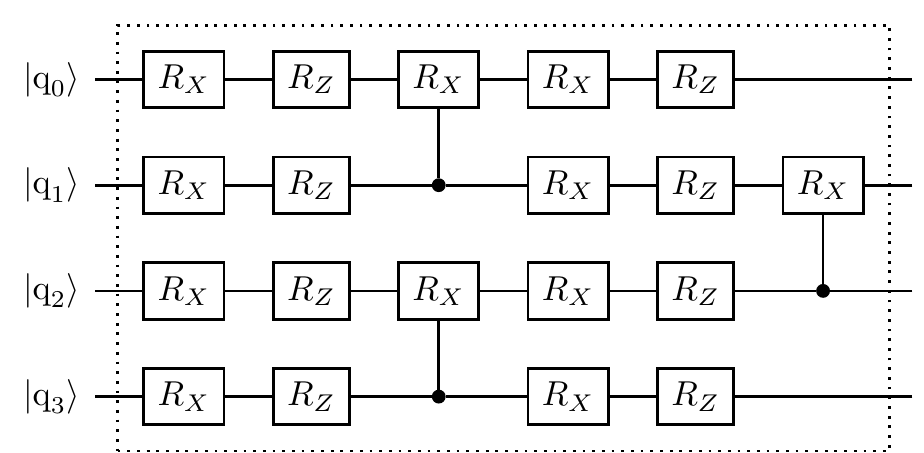}
\label{fig:circ-8}}
\subfloat[Circuit 11]{
\includegraphics[height=0.13\textwidth]{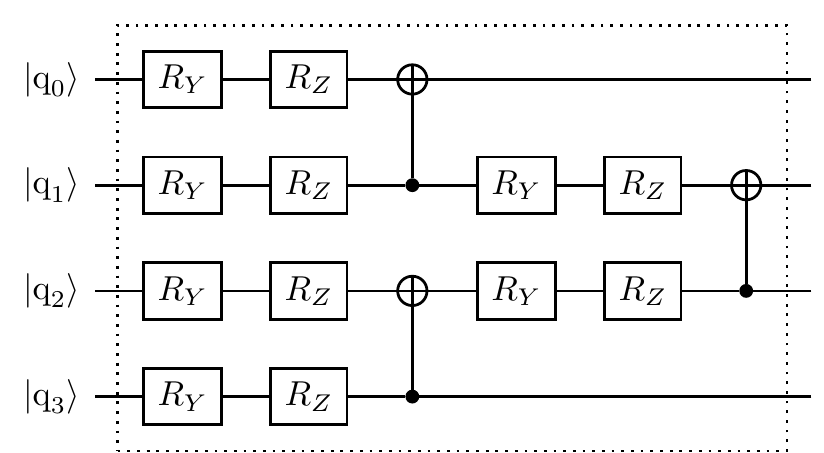}
\label{fig:circ-11}}
\subfloat[Circuit 12]{
\includegraphics[height=0.13\textwidth]{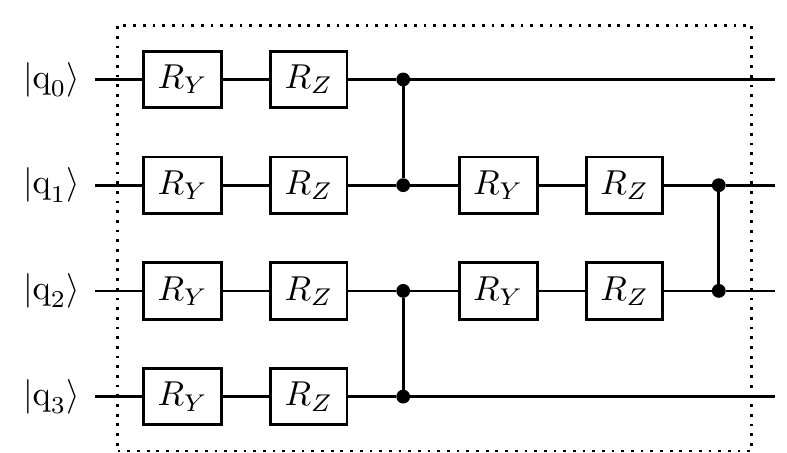}
\label{fig:circ-12}}\\~\\~\\
\subfloat[Circuit 13]{
\includegraphics[height=0.13\textwidth]{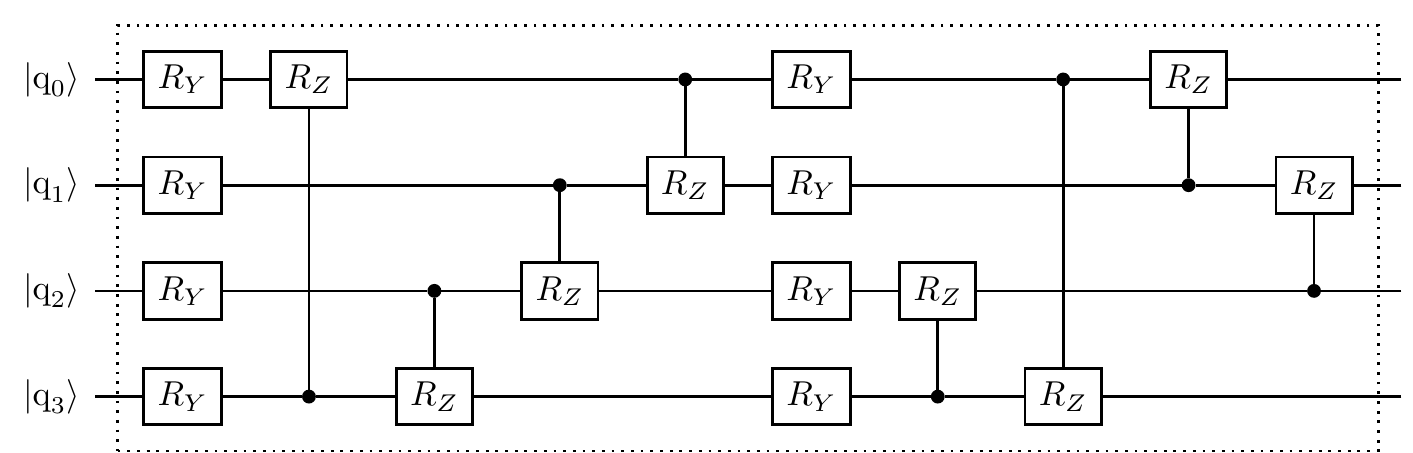}
\label{fig:circ-13}}
\subfloat[Circuit 14]{
\includegraphics[height=0.13\textwidth]{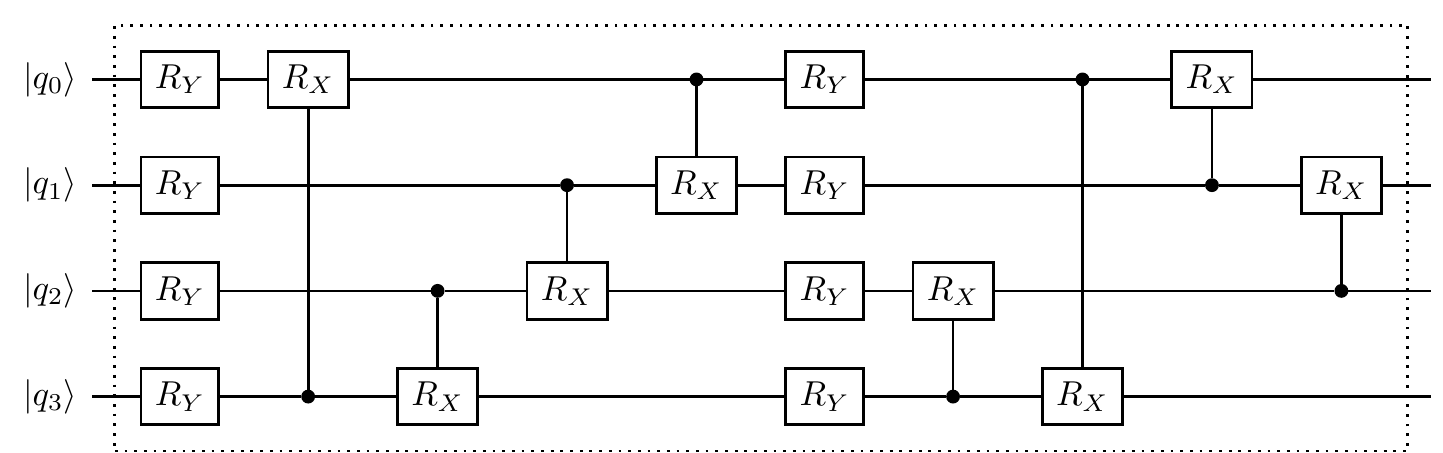}
\label{fig:circ-14}}\\~\\~\\
\subfloat[Circuit 15]{
\includegraphics[height=0.13\textwidth]{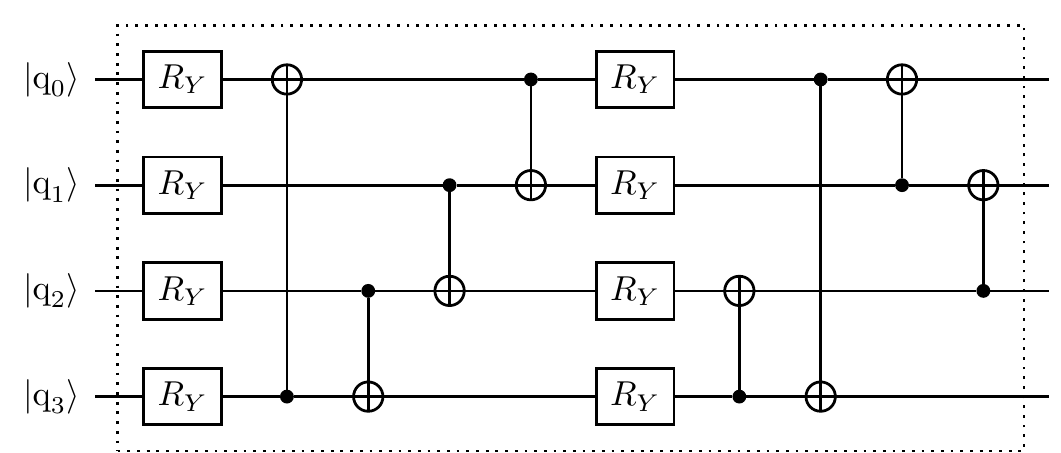}
\label{fig:circ-15}}
\subfloat[Circuit 16]{
\includegraphics[height=0.13\textwidth]{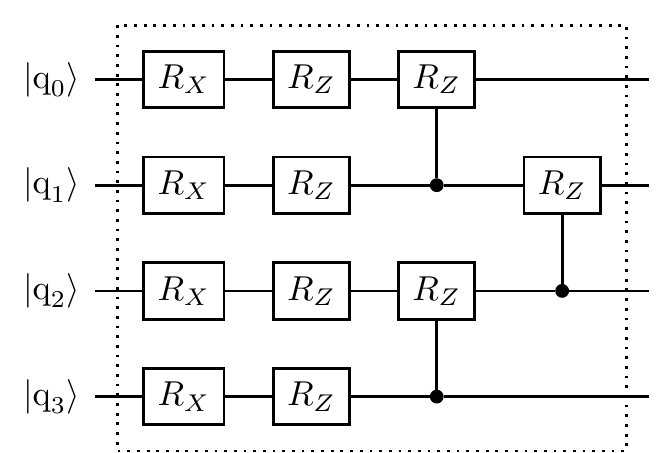}
\label{fig:circ-16}}
\subfloat[Circuit 17]{
\includegraphics[height=0.13\textwidth]{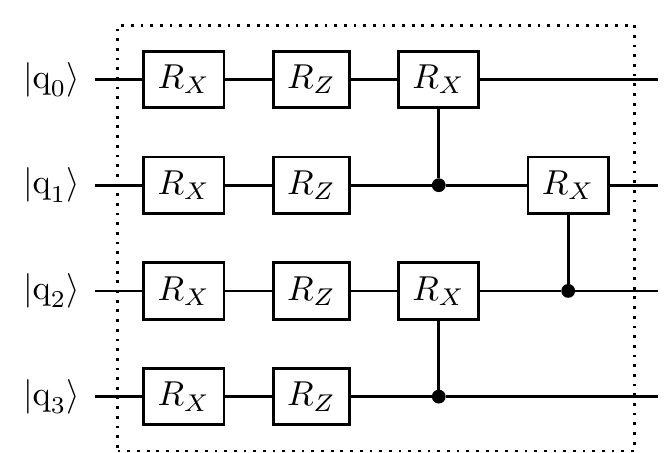}
\label{fig:circ-17}}
\subfloat[Circuit 18]{
\includegraphics[height=0.13\textwidth]{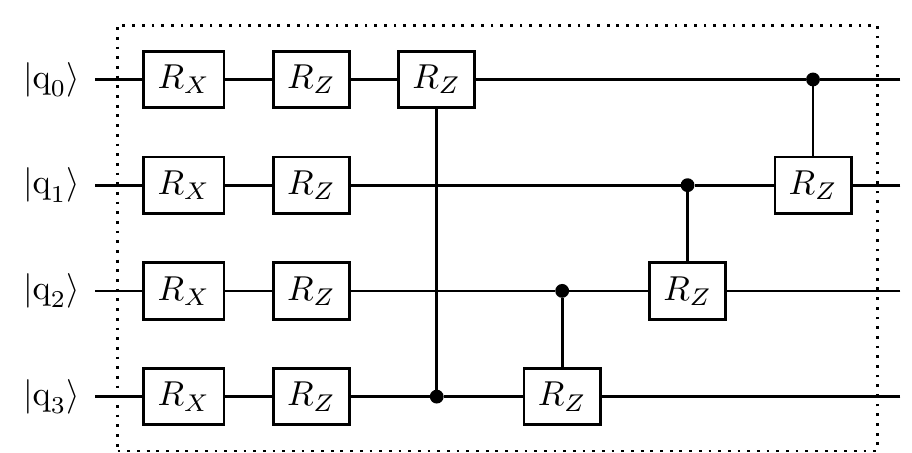}
\label{fig:circ-18}}\\~\\~\\
\subfloat[Circuit 19]{
\includegraphics[height=0.13\textwidth]{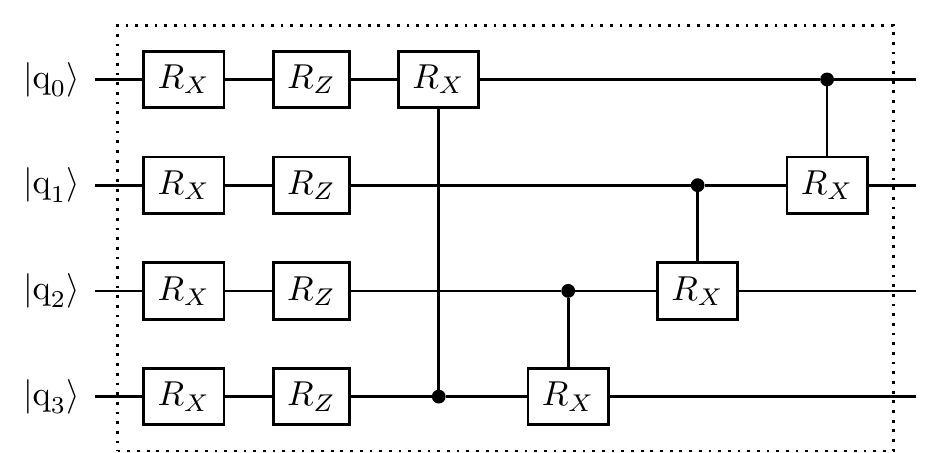}
\label{fig:circ-19}}\\~\\~\\
\caption{All 19 benchmarking quantum circuits, as presented in~\cite{sim2019expressibility}. These are the trainable part of the quantum kernel in the HQC RL solution. The input of each of these circuits is the output state of the basis embedding that encodes the state of the environment. }
\label{fig:all-circuits}
\end{figure*}

The circuit architectures are taken from the 19 circuits proposed in~\cite{sim2019expressibility} and are shown in Fig. \ref{fig:all-circuits}. An argument put forward by the authors is that these circuits were already designed for and successfully applied to diverse tasks. Moreover, all circuits are designed to be implementable on presently or soon available Noisy Intermediate-Scale Quantum (NISQ) devices, at various implementation costs. Four different entanglement gates are used: CRX, CRZ, CNOT, and CZ and five different entanglement topologies: linear, all-to-all, pairwise, circular, and shifted-circular-alternating. These categories are understood
as they are defined in the Qiskit documentation \cite{Qiskit}.

Circuit 1 is a basic quantum circuit with no entanglement and two degrees of freedom, with rotations around the X and Z axes of the Bloch sphere for each qubit. Circuit 2, 3, and 4 were designed by adding a basic linear entanglement using 3 different entanglement gates, with the purpose of studying the variation of the values of the quantum metrics for each type of entanglement gate and with respect to the first circuit. Circuits 5 and 6 were introduced as programmable universal quantum circuits in \cite{sousa2006universal} and used as quantum autoencoders in \cite{romero2017quantum}. Circuits 7 and 8 are part of the QVECTOR algorithm for quantum error correction \cite{johnson2017qvector} and circuit 9 was introduced as a "Quantum Kitchen Sinks" quantum machine learning architecture to be used on NISQ devices in \cite{wilson2018quantum}. The tenth circuit is taken from a hardware-efficient quantum architecture introduced in \cite{kandala2017hardware} as a variational quantum eigensolver, used to find the ground state energy for molecules. Circuits 11 and 12 are Josephson samplers defined in \cite{geller2018sampling} , whose purpose is to embed a vector of real elements into an $n$-qubits entangled state. Finally, circuits 13, 14, 15, 18, and 19 were constructed based on the generic model circuit architecture for classification tasks described in \cite{schuld2020circuit}. Circuit 16 and 17 are derived from circuits 3 and 4, but with the order of the last two controlled entanglement gates swapped. The purpose was to display the different expressibility values of circuits 3 and 16 and circuits 4 and 17 and emphasise that not only the type, but also the position of the quantum gates is important in a PQC.

After the trainable PQCs acted on $\psi_s$, the expectation value of the states are measured in the computational basis. The results are fed into a 4x4 linear layer in the case of the policy function. This linear layer outputs logits, which when normalised become the probabilities to choose an action between (left, down, right, up). This is thus the output of the policy function $\pi$. In the case of the state value function, the layer post-processing the expectation values has a 4x1 dimension to output only one value for an input state and to thus successfully model the value function approximator $V$. The loss is computed from the $\pi$ and $V$ function values and then used to classically adjust the parameters of the two PQCs accordingly. This pipeline is shown in Fig. \ref{fig:pipeline}.

%% file: chapters/05_results.tex
\section{\MakeUppercase{Performance of HQC Models}}
\label{sec:results}

For each classical and HQC architecture, three experiments were executed. The reward was sampled every 1000 time steps, and every experiment was ran for a total of 50000 time steps. These results were smoothed by a moving average over 10 data points. The error bands are calculated using the standard error of the reward achieved at a particular time step across all three experiments for a given architecture. No hyperparameter tuning was performed in this study due to the prohibitively long training times.

The RL metrics characterizing the performance of each solution variant are the maximum reward (MR) achieved during training and the time to convergence (TTC). The time to convergence is defined as the time step when the reward stabilises around its highest point. It is calculated as the first time step where a reward is reached that is not modified by more than 0.2 for all future time steps.

\begin{figure}[!h]
  \centering
    \includegraphics[width=0.475\textwidth]{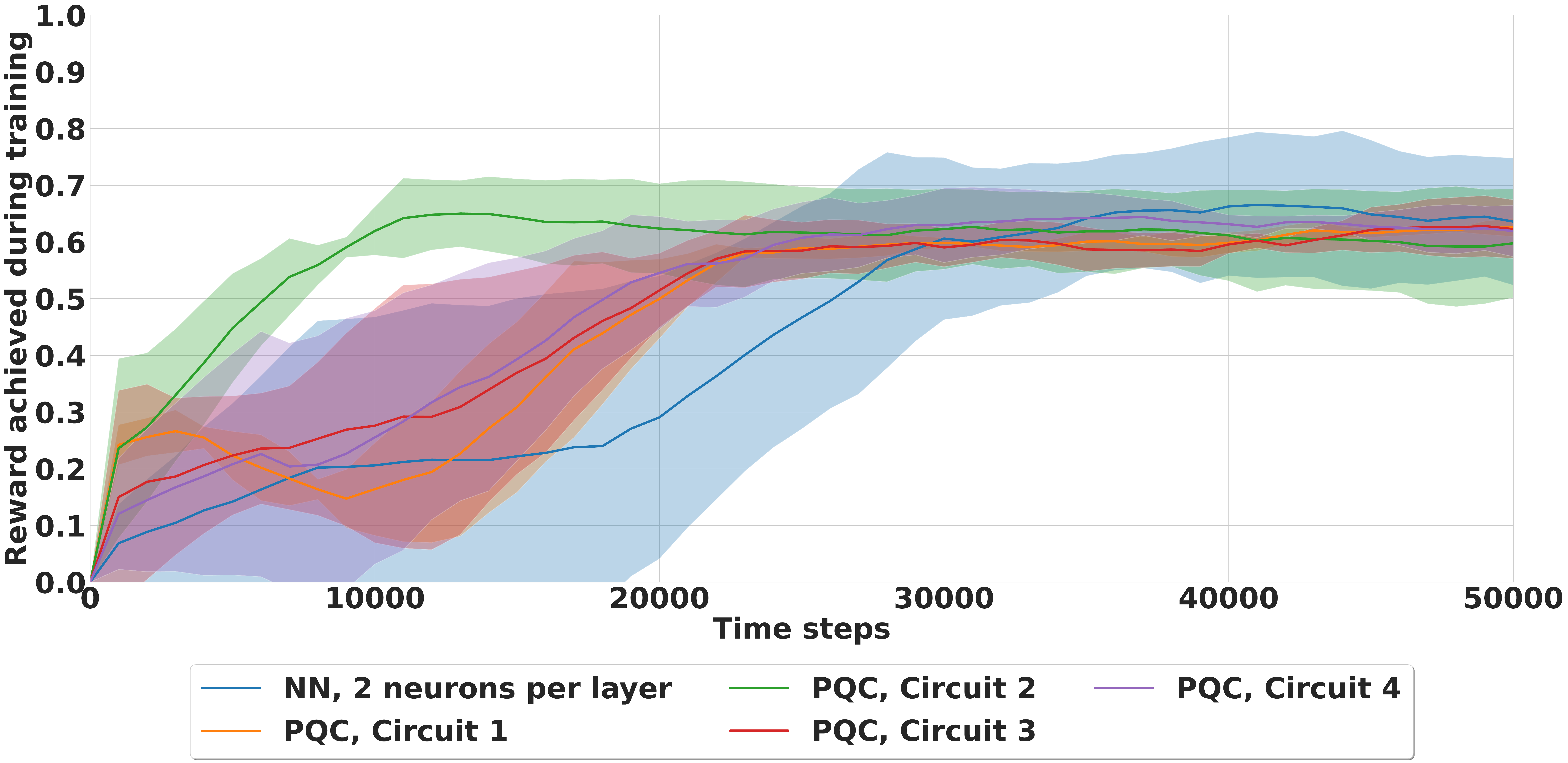}
  \caption{Comparison between the classical NN-based RL solution and the HQC solutions using PQCs 1 to 4 respectively, smoothed using a moving average.}
  \label{fig:NN-vs-PQC-1-small}
\end{figure}

\begin{figure}[!h]
  \centering
    \includegraphics[width=0.475\textwidth]{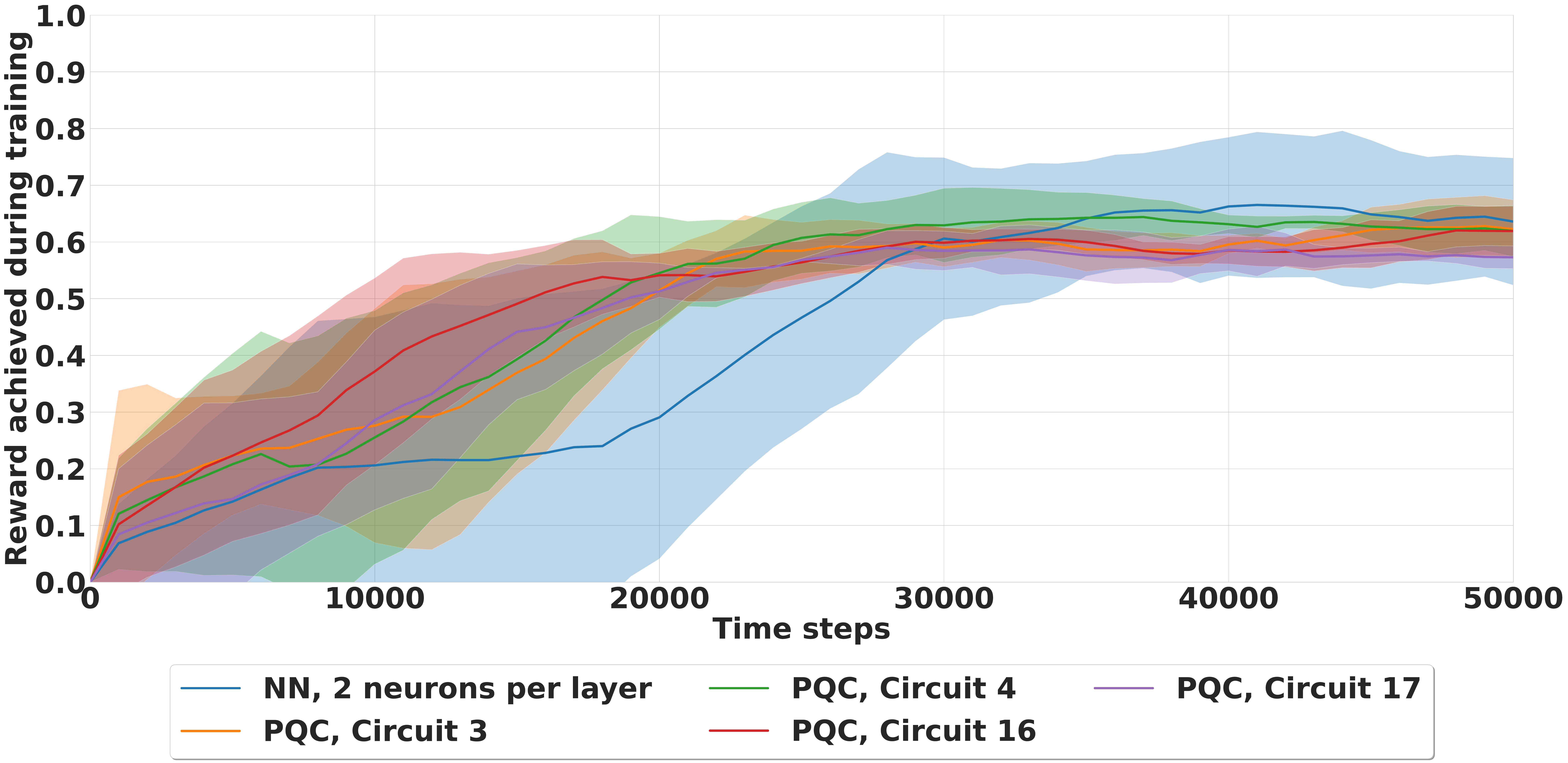}
  \caption{Comparison between the classical NN-based RL solution and the HQC solutions using PQCs 3, 4, 16, and 17, smoothed using a moving average.}
  \label{fig:NN-vs-PQC-2-small}
\end{figure}

\begin{figure}[!h]
  \centering
    \includegraphics[width=0.475\textwidth]{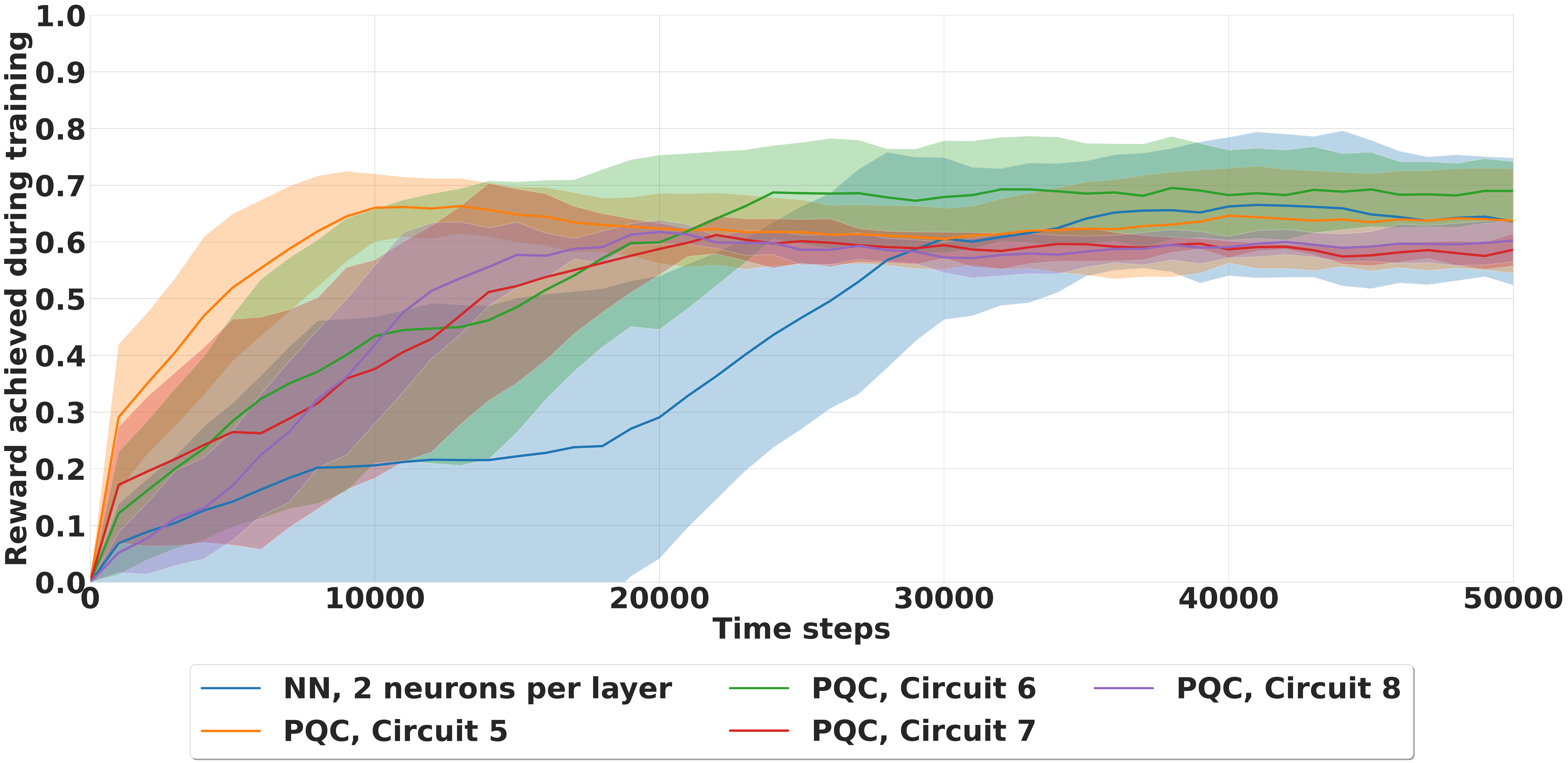}
  \caption{Comparison between the classical NN-based RL solution and the HQC solutions using PQCs 5 to 8 respectively, smoothed using a moving average.}
  \label{fig:NN-vs-PQC-3-small}
\end{figure}

\begin{figure}[!h]
  \centering
    \includegraphics[width=0.475\textwidth]{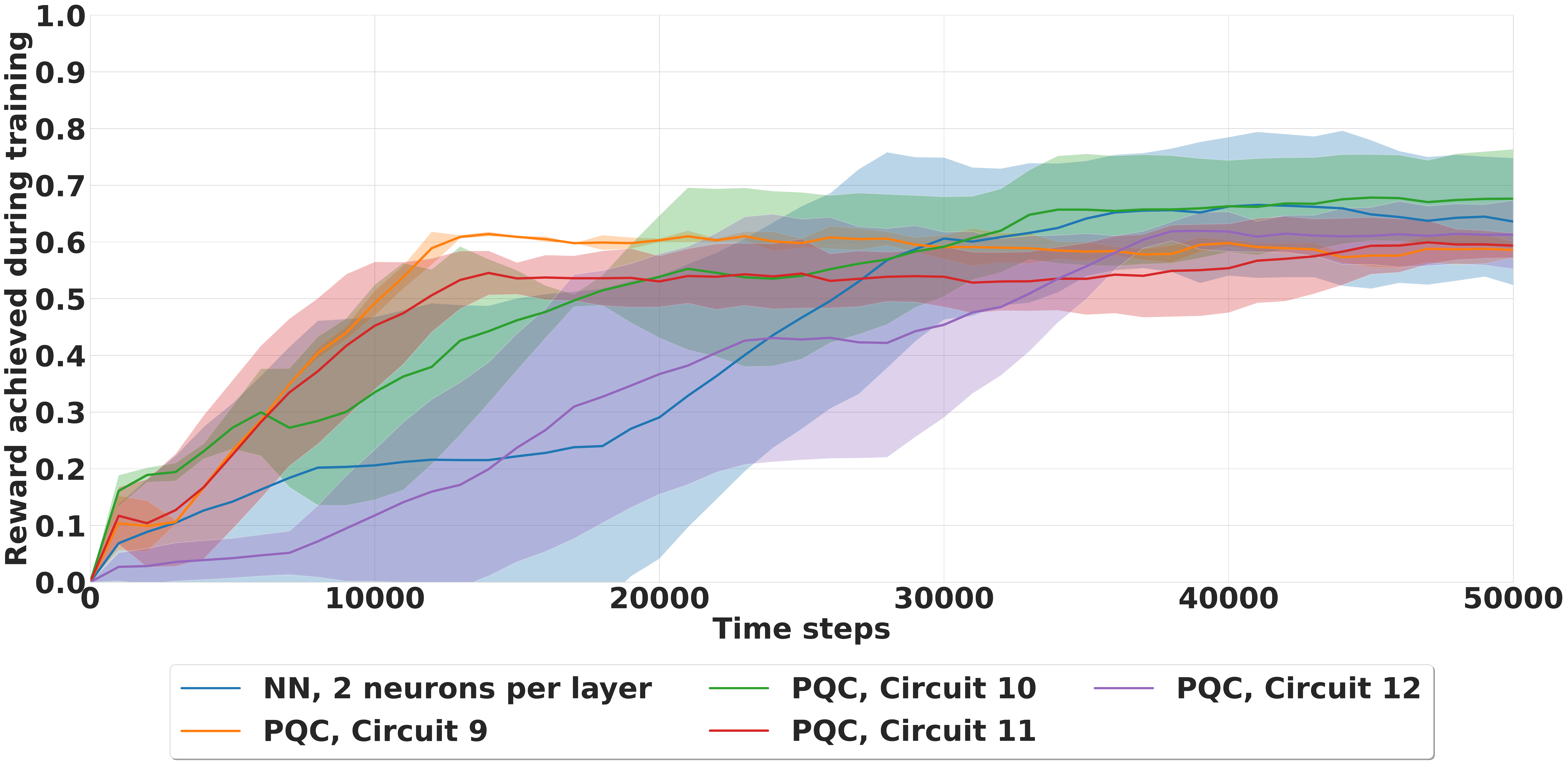}
  \caption{Comparison between the classical NN-based RL solution and the HQC solutions using PQCs 9 to 12 respectively, smoothed using a moving average.}
  \label{fig:NN-vs-PQC-4-small}
\end{figure}

\begin{figure}[!h]
  \centering
    \includegraphics[width=0.475\textwidth]{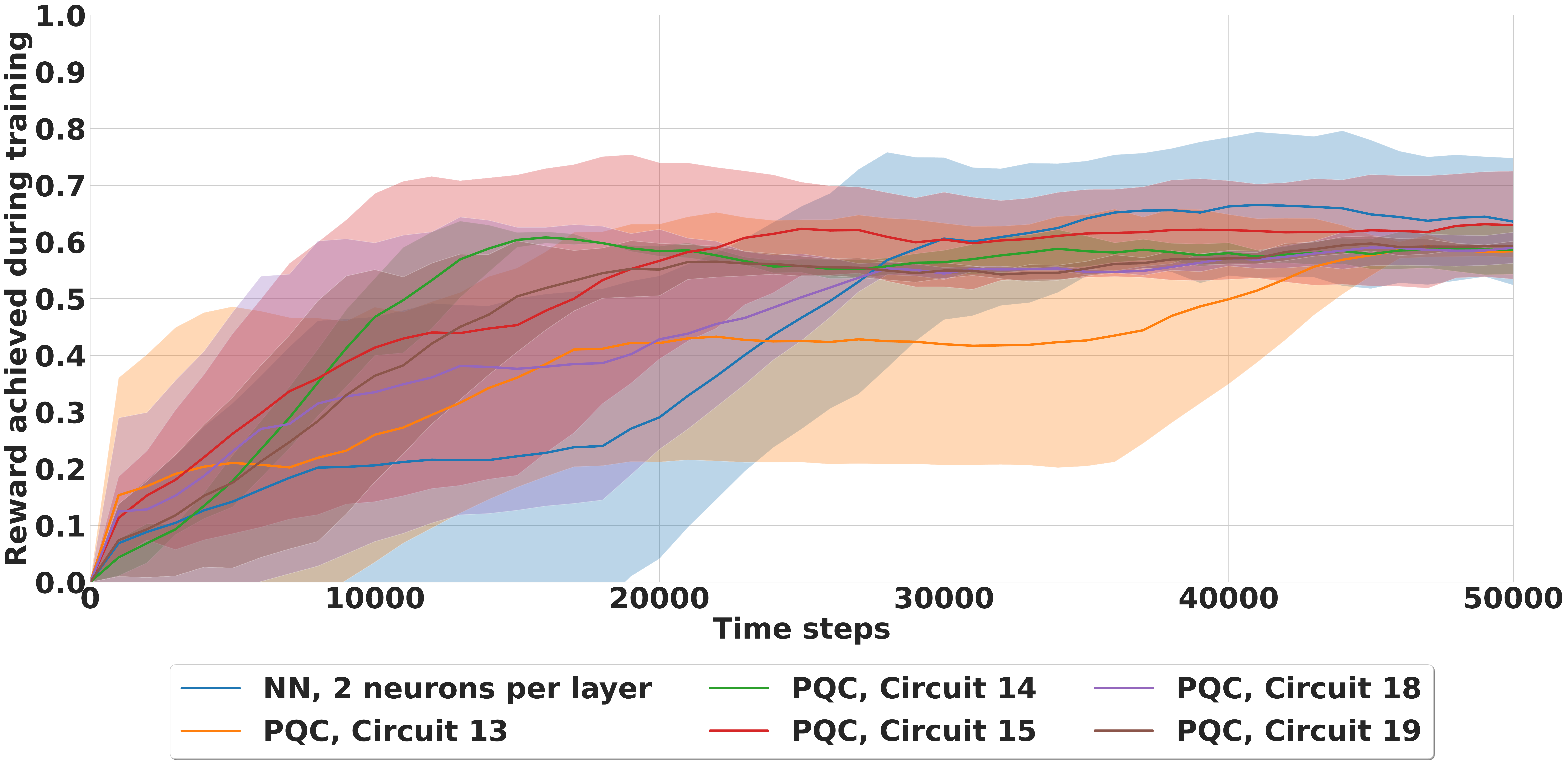}
  \caption{Comparison between the classical NN-based RL solution and the HQC solutions using PQCs 13 to 19 respectively, smoothed using a moving average.}
  \label{fig:NN-vs-PQC-5-small}
\end{figure}

The rewards for all quantum circuits are plotted in Figs. \ref{fig:NN-vs-PQC-1-small} to \ref{fig:NN-vs-PQC-5-small} and detailed in Table \ref{table:results_all}. In general, the HQC solutions and the classical solutions achieved around the same average MR: the best value was 0.86 for the classical approach using four neurons per hidden layer and 0.85 for the HQC algorithm when considering circuit 6 from Fig.~\ref{fig:all-circuits}. Both types of solutions were successful and passed the computed learning threshold of $\text{r}_{\text{thr}} = 0.81$.
Nevertheless, when also considering the number of trainable parameters, one can see that the HQC approach only needs a third of the number of trainable parameters necessary for the classical one to perform comparatively well -- 81 compared to 237 trainable parameters. Moreover, the TTC is clearly smaller for most of the variants of HQC variants tested, as shown in Figs. \ref{fig:NN-vs-PQC-1-small} -- \ref{fig:NN-vs-PQC-5-small}. The lowest mean TTC is 10330 time steps for the HQC solution of circuit 2 with 41 trainable parameters. In order to achieve stabilization that fast, a classical solution would require 1245 trainable parameters. Generally, the best HQC solutions were the ones employing circuits 2, 6, and 10. Circuits 7-19 generally learn faster than the classical solution with a much higher reward in the first 25000 time steps, but stabilise afterwards at a lower reward in our tests.

\begin{table*}[!h]
    \centering
    \caption{The results of all architectures used to solve the FL, evaluated with respect to the MR and the TTC (in thousands of time steps) obtained during training. The MR and TTC values are obtained from the initial training processes, before the smoothing that leads to Figs. \ref{fig:NN-vs-PQC-1-small} to \ref{fig:NN-vs-PQC-5-small}. We also present the number of the trainable weights of each solution (W), together with their expressibility (Exp), entanglement capability (Ent), and the effective dimension (ED). The results below are in ascending order from the best to the worst values of the TTC and are grouped into classical and HQC solutions. For the quantum metrics, the higher the Ent and ED values, and the lower the Exp values, the better the circuit is in reference to the respective metric. The degree of correlation between the number of weights W and the MR and TTC of each solution remains for now unclear.}
    \begin{tabular}[width=\textwidth]{| l | c | c | c | c | c | c | c | c |}
       \hline
        Solution & W & MR & TTC & Ent & Exp & ED \\
        \hline
        PQC--2 & 41 & 0.77 $\pm$ 0.16 & 10.33 $\pm$ \hphantom{0}7.58 & 0.81 & 0.28 & 3.50 \\ 
        PQC--5 & 81 & 0.78 $\pm$ 0.22 & 11.33 $\pm$ \hphantom{0}3.79 & 0.41 & 0.06 & 6.91 \\ 
        PQC--11 & 49 & 0.71 $\pm$ 0.20 & 12.33 $\pm$ \hphantom{0}5.17 & 0.73 & 0.13 & 5.08 \\ 
        PQC--9 & 33 & 0.75 $\pm$ 0.02 & 12.50 $\pm$ \hphantom{0}1.24 & 1.00 & 0.67 & 3.48 \\ 
        PQC--8 & 63 & 0.72 $\pm$ 0.08 & 14.33 $\pm$ 10.34 & 0.39 & 0.08 & 6.24 \\ 
        PQC--19 & 49 & 0.71 $\pm$ 0.07 & 14.33 $\pm$ 12.50 & 0.59 & 0.08 & 6.29 \\ 
        PQC--7 & 63 & 0.72 $\pm$ 0.06 & 15.33 $\pm$ 16.16 & 0.33 & 0.09 & 5.82 \\ 
        PQC--14 & 57 & 0.78 $\pm$ 0.25 & 16.33 $\pm$ \hphantom{0}7.58 & 0.66 & 0.01 & 7.68 \\ 
        PQC--15 & 41 & 0.76 $\pm$ 0.28 & 19.67 $\pm$ 16.54 & 0.82 & 0.19 & 4.60 \\ 
        PQC--16 & 47 & 0.78 $\pm$ 0.09 & 20.00 $\pm$ 25.21 & 0.35 & 0.26 & 3.73 \\ 
        PQC--18 & 49 & 0.72 $\pm$ 0.10 & 20.00 $\pm$ 26.28 & 0.44 & 0.23 & 3.70 \\ 
        PQC--1 & 41 & 0.72 $\pm$ 0.10 & 21.67 $\pm$ 10.34 & 0.00 & 0.29 & 3.29 \\ 
        PQC--4 & 47 & 0.81 $\pm$ 0.18 & 23.67 $\pm$ 14.12 & 0.47 & 0.13 & 5.58 \\ 
        PQC--17 & 47 & 0.72 $\pm$ 0.09 & 25.00 $\pm$ \hphantom{0}9.93 & 0.4 & 0.13 & 5.74 \\ 
        PQC--13 & 57 & 0.72 $\pm$ 0.08 & 25.00 $\pm$ 53.79 & 0.61 & 0.05 & 7.07 \\ 
        PQC--6 & 81 & 0.85 $\pm$ 0.16 & 26.00 $\pm$ \hphantom{0}4.96 & 0.78 & 0.00 & 7.79 \\ 
        PQC--12 & 49 & 0.75 $\pm$ 0.19 & 26.66 $\pm$ 27.92 & 0.65 & 0.20 & 4.91 \\ 
        PQC--3 & 47 & 0.79 $\pm$ 0.06 & 27.67 $\pm$ 48.25 & 0.34 & 0.24 & 3.72 \\ 
        PQC--10 & 41 & 0.81 $\pm$ 0.27 & 31.67 $\pm$ 14.34 & 0.54 & 0.22 & 3.98 \\ 
        \hline
        NN--16 & 1245 & 0.81 $\pm$ 0.10 & 11.33 $\pm$ \hphantom{0}3.12 & -- & -- & 48.78 \\ 
        NN--2 & 125 & 0.84 $\pm$ 0.04 & 19.00 $\pm$ 15.51 & -- & -- & 42.53 \\ 
        NN--4 & 237 & 0.86 $\pm$ 0.00 & 22.00 $\pm$ \hphantom{0}5.61 & -- & -- & 72.13 \\ 
        NN--8 & 509 & 0.85 $\pm$ 0.02 & 24.33 $\pm$ \hphantom{0}6.84 & -- & -- & 74.83 \\ 
        \hline
    \end{tabular}
    \label{table:results_all}
\end{table*}

%% file: chapters/03_quantum_metrics.tex
\section{\MakeUppercase{The dependence of performance on quantum metrics}}
\label{sec:quantum_metrics}

The circuits proposed in~\cite{sim2019expressibility} were designed for different purposes and indeed show distinct performances in our tests. Different metrics are discussed in the literature to describe the power of specific quantum circuits -- comparing to other quantum circuits, but in part also allow the comparison to alternative classical implementations. It is however unclear from the literature if these metrics are correlated with the performance of a QRL model or not. Therefore, currently indications how a quantum kernel should be constructed inside a QRL model are missing. There are in particular three metrics that are commonly discussed to characterise the properties of quantum circuits:

\begin{enumerate}
    \item \textbf{Expressibility}: This metric measures how well a PQC covers the entire Hilbert space and therefore indicates if the model would theoretically be able to learn a target function.
    \item \textbf{Entanglement capability}: The power of using quantum circuits in comparison to classical variants also arises from the possibility to entangle qubits, and to therefore strongly correlate them. One may thus wonder if a higher entanglement within a PQC leads to an improved performance of the final model.
    \item \textbf{Effective dimension:} The effective dimension allows a direct comparison of classical NN and quantum neural networks (QNN) (i.e., PQC) by interpreting both of them as statistical models where one can measure the information capacity. It is known \cite{abbas2021power} that specific QNNs exhibit a higher effective dimension than equivalent classical models and that the concerned QNNs may show a better performance .
\end{enumerate}

In the following, we briefly describe the mathematical definitions of these metrics. Afterwards, we discuss their correlation with the performance of our QRL models.

\subsection{Expressibility}
The expressibility of a circuit \cite{sim2019expressibility} measures how well a set of (pure) states, here generated by the PQC, covers the entire Hilbert space. The precise calculation is described in~\cite{sim2019expressibility}. The calculation uses the Kullback-Leibler (KL) divergence \cite{kullback1951information} between two distributions of fidelities. These belong to the Haar ensemble of random states and respectively to the ensemble of states generated by the PQC. After having chosen a PQC architecture, one uniformly samples two parameter vectors $\theta_i$ and $\theta_j$ from the entire parameter space of the PQC. Then the fidelity of their corresponding states $\ket{\psi_i}$ and $\ket{\psi_j}$ is computed. This process is repeated multiple times and the results are afterwards plotted as a histogram of the probability density function. For the Haar ensemble, the analytic probability density function of fidelities is
\begin{equation}
P_{\text{Haar}} = (N-1)(1-F)^{(N-2)},
\end{equation}
where $F$ is the fidelity and $N$ is the dimension of the Hilbert space. Finally, the KL divergence is calculated from the two histograms and the result is the $D_{KL}$ divergence, whose value is inversely proportional to how expressive the circuit is:
\begin{equation}
    \text{Exp} = D_{\text{KL}}(P_{\text{PQC}}(F;\theta)~||~P_{\text{Haar}}(F)).
\end{equation}
\subsection{Entanglement Capability}
In order to quantify the degree of entanglement, the authors of \cite{sim2019expressibility} employed the Meyer-Wallach (MW) entanglement measure \cite{meyer2002global}. The equation to assess the entanglement capability is 
\begin{equation}
    \text{Ent} = \frac{1}{|S|} \sum_{\theta_i \in S} Q(\ket{\psi_i}),
\end{equation}
where $S = \{ \theta_i \}$ is the set of sampled parameter vectors $\theta_i$, and Q is the MW measure. It is defined as 
\begin{equation}
    Q(\ket{\psi}) \doteq \frac{4}{n} \sum_{j=1}^n D(\iota_j(0) \ket{\psi}, \iota_j(1) \ket{\psi}),
\end{equation}
where $n$ is the number of qubits in the system and $D$ is the generalised distance:
\begin{equation}
    D(\ket{u}, \ket{v}) = \frac{1}{2} \sum_{i, j} |u_i v_j - u_j v_i|^2.
\end{equation}
The linear mapping $\iota_j(b)$ acts on a quantum basis state:
\begin{equation}
    \iota_j(b) \ket{b_1\dots b_n} \doteq \delta_{bb_j} \ket{b_1\dots b_{j-1}  b_{j+1} \dots b_n},
\end{equation}
where the $b_j$ qubit disappears and $\delta$ is the Kronecker-Delta operator.

\subsection{Effective Dimension}
The Effective Dimension (ED) characterises both classical and quantum machine learning models in terms of their power to generalise and fit to the given data. It is derived from the Fisher Information Matrix (FIM), which is a metric in statistics that assesses the impact of the variance of the parameters of the model on its output. In our case, it is computed using the probability $p(x,~y;~\theta)$, which shows the relationship between the input $x \in \mathbb{R}^{s_{in}}$, the output $y \in \mathbb{R}^{s_{out}}$, and the parameters $\theta$. The Riemannian space of the model parameters is $\Theta \subset \mathbb{R}^d$. The FIM is then computed as 
\begin{equation}
    F(\theta) = \mathbb{E} \left[ \frac{\partial}{\partial \theta} \log p(x,~y;~\theta)~ \frac{\partial}{\partial \theta} \log p(x,~y;~\theta)^T \right],
\end{equation}
where $F \in \mathbb{R}^{d \times d}$. If one uses finite sampling, the FIM can be approximated empirically to
\begin{equation}
    \Tilde{F}_k(\theta) = \frac{1}{k} \sum_{j=1}^k \frac{\partial}{\partial \theta} \log p(x_j,~y_j;~\theta)~ \frac{\partial}{\partial \theta} \log p(x_j,~y_j;~\theta)^T,
\end{equation}
where $k$ is the number of independent and identically distributed (i.i.d) samples $(x_j, y_j)$ drawn from $p(x,~y;~\theta)$. The formulation of the ED is based on \cite{berezniuk2020scale}. The authors of \cite{abbas2021power} extend this by adding the constant $\gamma \in (0,1]$ and a $\log n$ term to assure the ED is bounded. This results in the final form we use in this work:
\begin{equation}
    d_{\gamma, n}(\mathcal{M}_\Theta) = \frac{\log(\frac{1}{V_\Theta} \int_\Theta \sqrt{ \det (\text{id}_d + \frac{\gamma n}{2 \pi \log n} \hat{F}(\theta))}~d\theta)}{\log(\frac{\gamma n}{2 \pi \log n})}, 
\end{equation}
where $n > 1 \in \mathbb{N}$ is the number of data samples,  $V_\Theta \doteq \int_\Theta d\theta$ is the volume of the parameter space, and $\hat{F}(\theta)$ is the normalised FIM formulated as:
\begin{equation}
    \hat{F}_{ij} (\theta)= d \frac{V_\Theta}{\int_\Theta tr(F(\theta))~d\theta} F_{ij}(\theta).
\end{equation}

\subsection{Impact of the Metrics on the Performance }

\begin{figure*}[!h]
  \centering
  \captionsetup{width=0.80\linewidth}
  \includegraphics[width=0.80\textwidth]{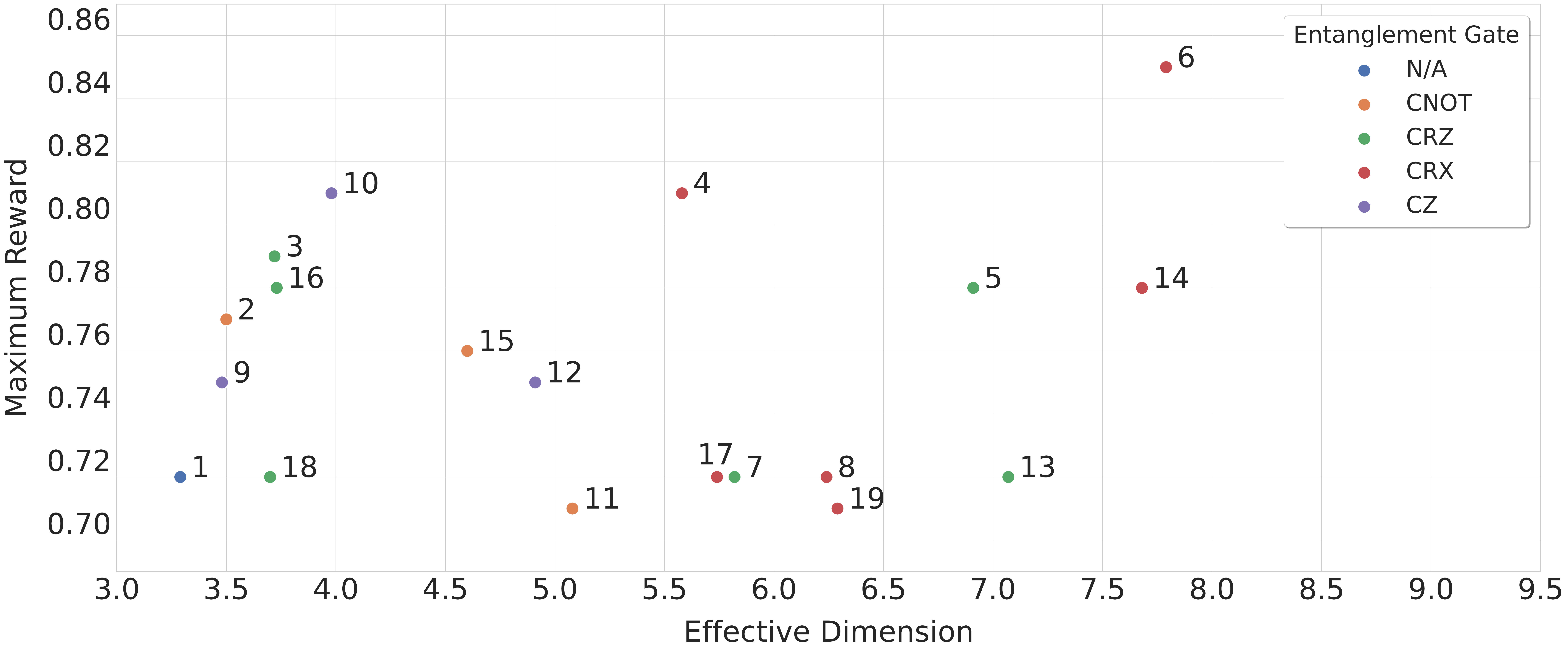}
  \caption{Correlation between the ED of a PQC and the MR obtained by the agent employing it, labeled using the PQC index and aggregated by the entanglement gate.}
  \label{fig:ED-vs-MR-EG}
\end{figure*}

\begin{figure*}[!h]
  \centering
  \captionsetup{width=0.80\linewidth}
  \includegraphics[width=0.80\textwidth]{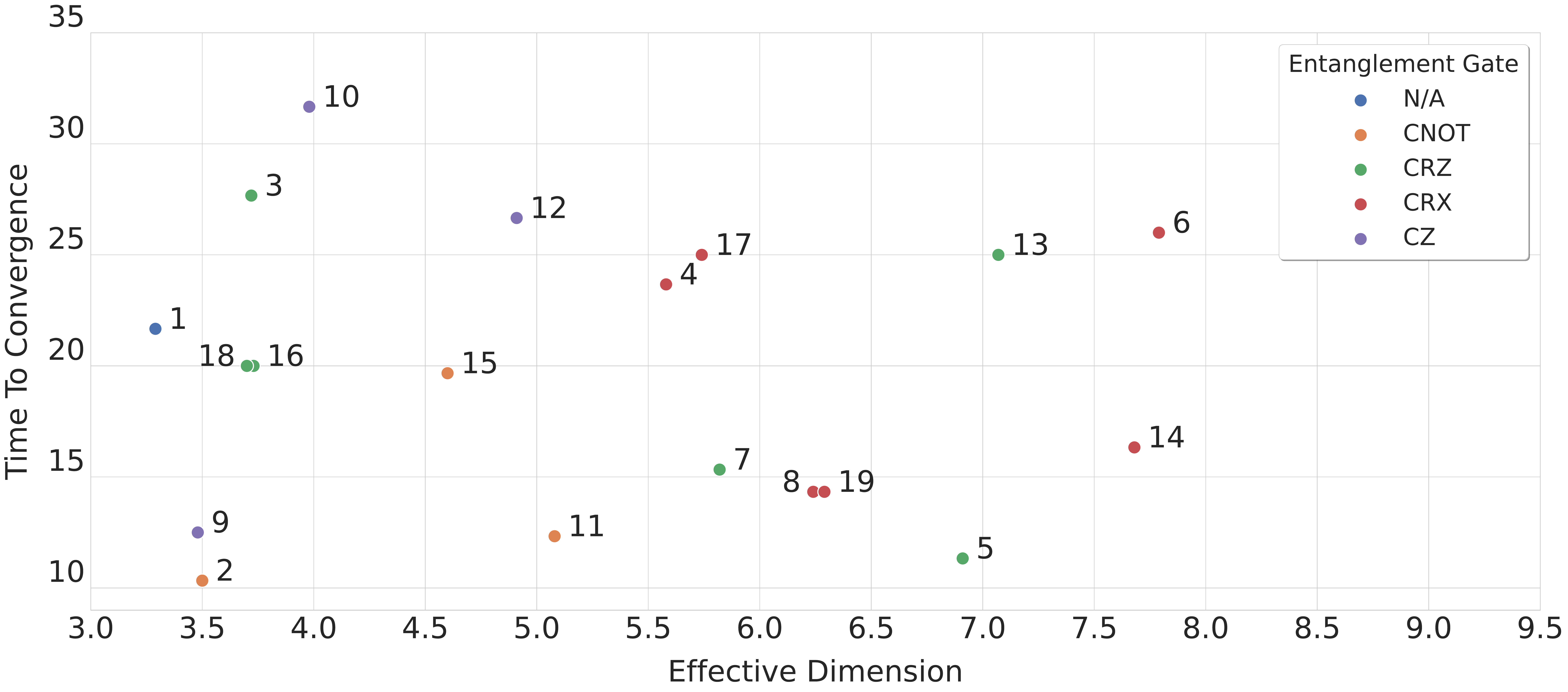}
  \caption{Correlation between the ED of a PQC and the TTC obtained by the agent employing it, labeled using the PQC index and aggregated by the entanglement gate.}
  \label{fig:ED-vs-TTC-ET}
\end{figure*}

\begin{figure*}[!h]
  \centering
  \captionsetup{width=0.80\linewidth}
  \includegraphics[width=0.80\textwidth]{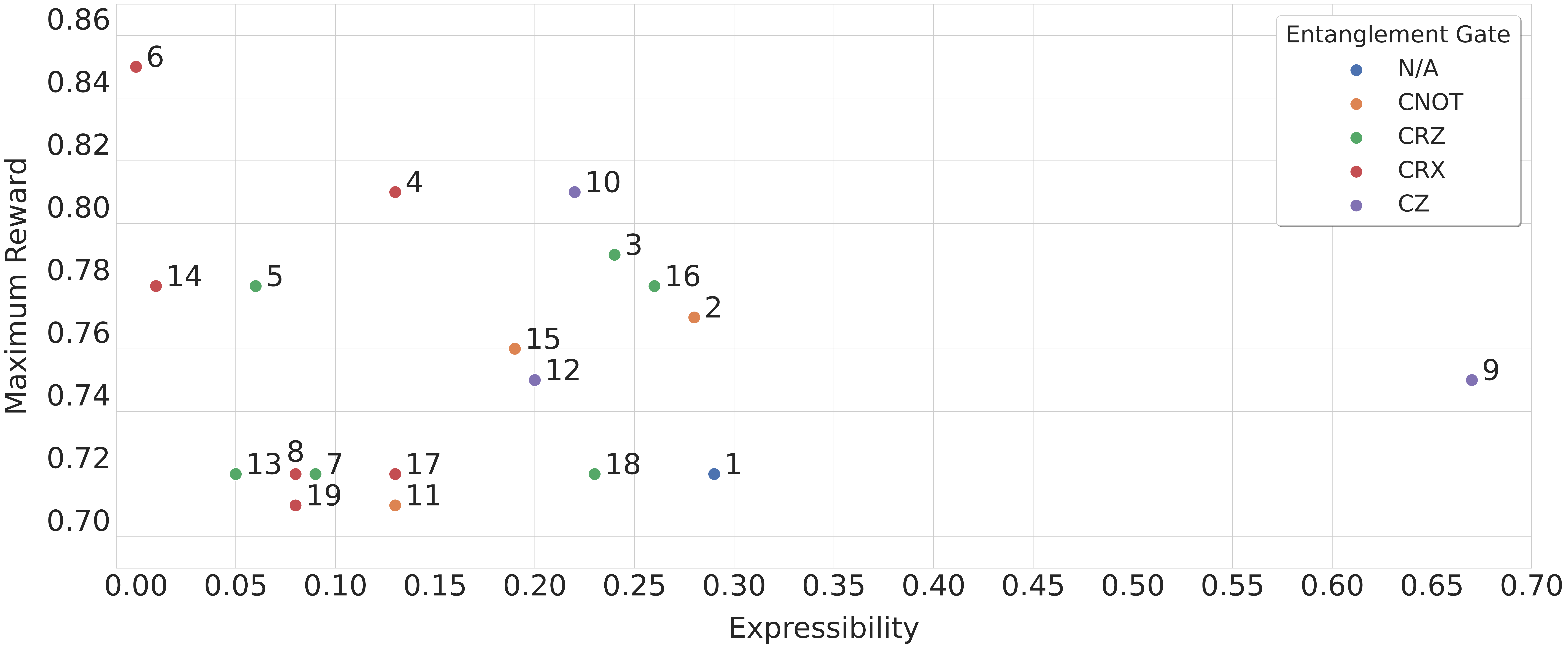}
  \caption{Correlation between the expressibility of a PQC and the MR obtained by the agent employing it, labeled using the PQC index and aggregated by the entanglement gate.}
  \label{fig:Expr-vs-MR-RY}
\end{figure*}

\begin{figure*}[!h]
  \centering
  \captionsetup{width=0.80\linewidth}
  \includegraphics[width=0.80\textwidth]{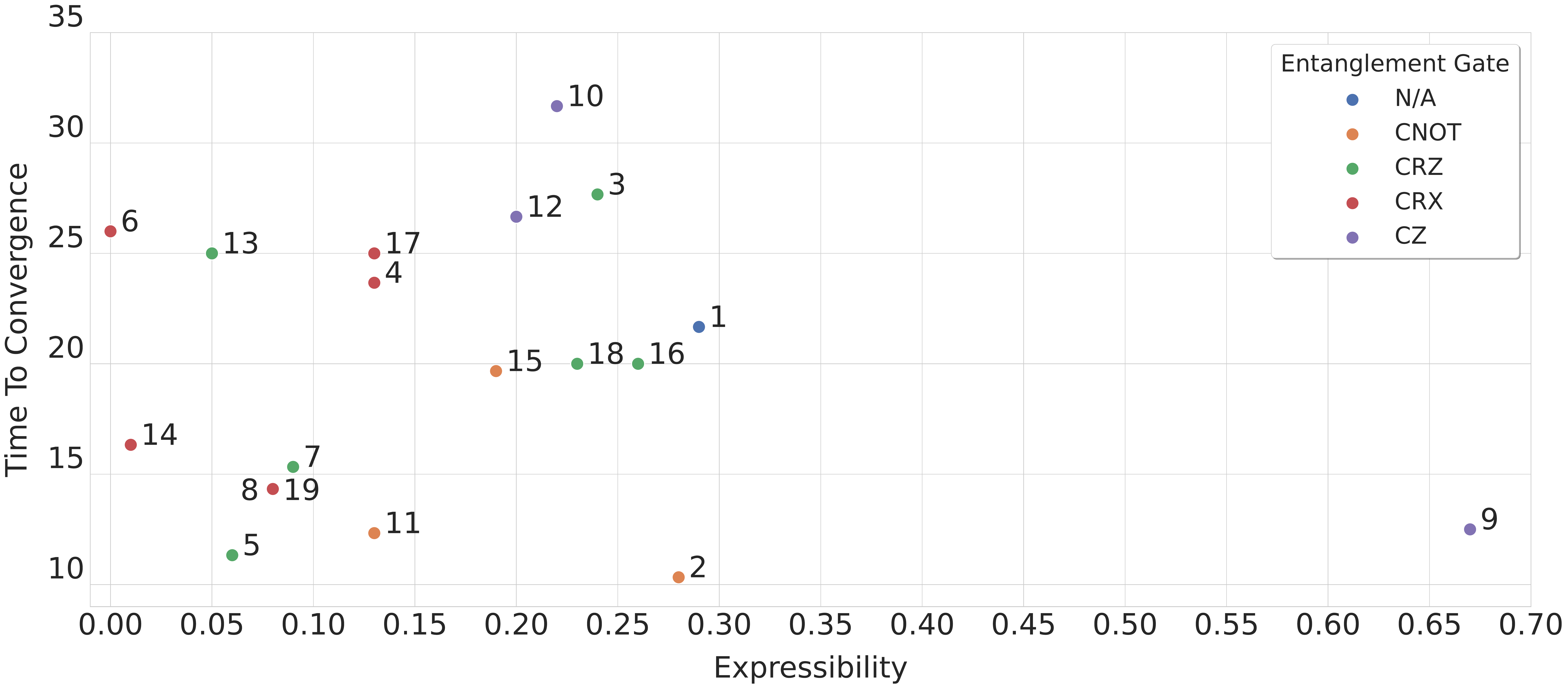}
  \caption{Correlation between the expressibility of a PQC and the TTC obtained by the agent employing it, labeled using the PQC index and aggregated by the entanglement gate.}
  \label{fig:Expr-vs-TTC-EG}
\end{figure*}

\begin{figure*}[!h]
  \centering
  \captionsetup{width=0.80\linewidth}
  \includegraphics[width=0.80\textwidth]{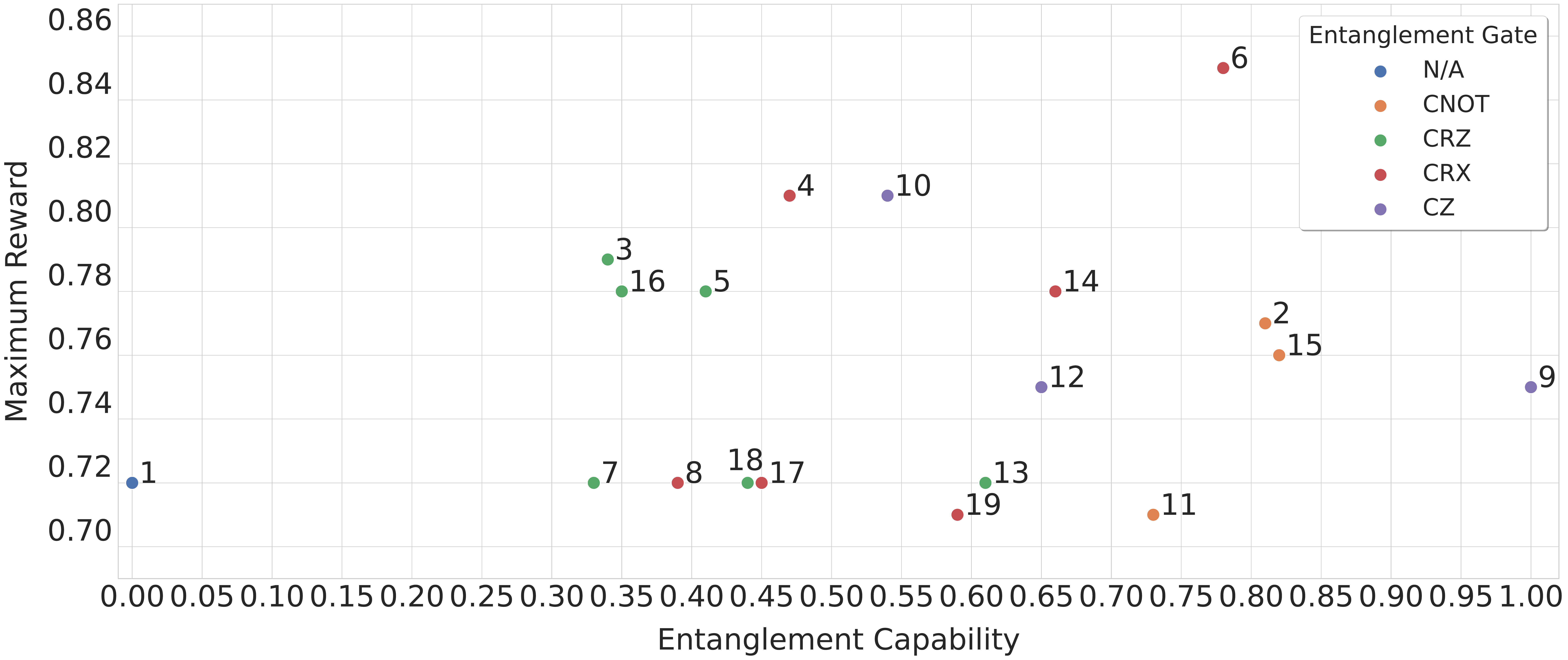}
  \caption{Correlation between the entanglement capability of a PQC and the MR obtained by the agent employing it, labeled using the PQC index and aggregated by the entanglement gate.}
  \label{fig:Ent-vs-MR-ET}
\end{figure*}

\begin{figure*}[!h]
  \centering
  \captionsetup{width=0.80\linewidth}
  \includegraphics[width=0.80\textwidth]{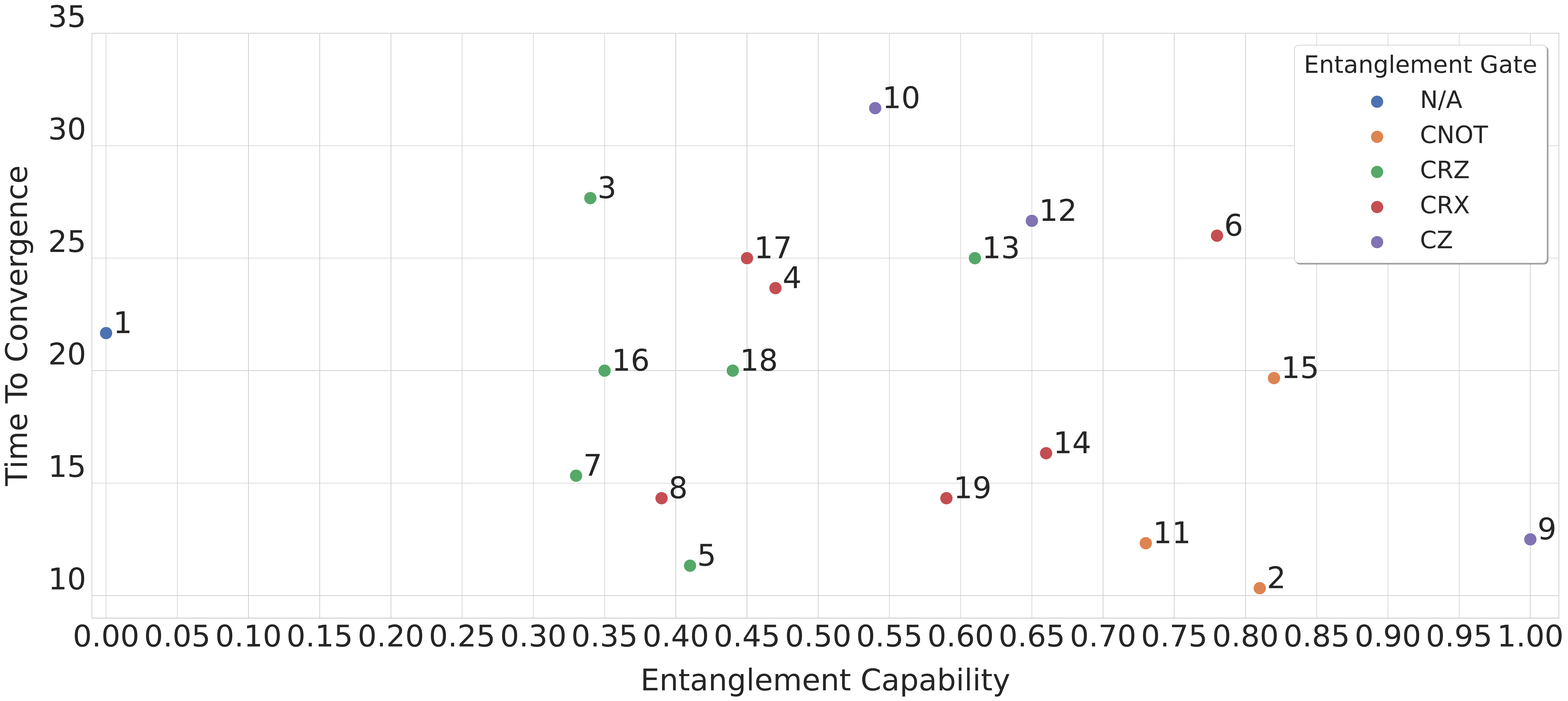}
  \caption{Correlation between the entanglement capability of a PQC and the TTC obtained by the agent employing it, labeled using the PQC index and aggregated by the entanglement gate.}
  \label{fig:Ent-vs-TTC-RY}
\end{figure*}

Values for the expressibility and the entanglement capability for all considered circuits can be directly taken from~\cite{sim2019expressibility}. The ED in contrast is calculated by us. These three metrics are displayed in Table \ref{table:results_all} alongside the MR and TTC values, as well as comparatively plotted in Figs. \ref{fig:ED-vs-MR-EG} to \ref{fig:Ent-vs-TTC-RY}.

We first observe that circuits 2, 3, and 4 largely overcome circuit 1 -- the circuit with no entanglement -- in MR, as expected. Similarly, circuit 2 performs significantly better than circuit 3 in the TTC (while having a similar MR). The expressibility and ED of the two circuits being comparable, this seems to indicate that the entanglement capacity plays a positive role in the final performance.
It is interesting to note that circuits 2 and 4 are very similar in structure, while circuit 4 containing three more trainable parameters than circuit 2. The last three gates perform rotations around the x-axis in case of both circuits. Despite this similarity, circuit 4 with more trainable parameters results in a worse performance in the TTC than circuit 2. 

When inspecting the performance of further circuits, we do not observe a clear correlation between the MR, the TTC, and the values of expressibility, entanglement capacity, and effective dimension, as shown in Figs. \ref{fig:ED-vs-MR-EG} to \ref{fig:Ent-vs-TTC-RY}. For example, the best performing circuit considering the average MR is circuit 6. It also has the best expressibility, the highest ED, and the fourth highest entanglement capability, which would seem to indicate a positive correlation. On the other hand, circuit 13, one of the worse performing circuits in MR, has a better expressibility, a higher entanglement capability, and a higher effective dimension than circuit 10, which is, along with circuit 6, one of the best performing circuits in MR.

In terms of the TTC, we find circuits supporting the hypothesis that better metrics lead to improved performance, such as circuit 2 and its large entanglement capacity; as well as counterexamples, such as circuit 5, which is the second fastest in TTC, but is overcome in all three metrics by circuit 6 (and with exactly the same number of trainable parameters). Additionally, in view of the dissimilar performances between circuits 3 and 16 as well as circuits 4 and 17, we deduce that the position of the entangling gates can impact the performance, which is a factor that is not considered in the three chosen metrics.

%% file: chapters/06_conclusions.tex
\section{CONCLUSIONS}
\label{sec:conclusions}

This work presents a hybrid quantum-classical reinforcement learning algorithm that successfully solves a stochastic slippery 4x4 frozen lake example with a probability of 80\% to move into the desired direction and 20\% to move into undesired orthogonal directions. The algorithms considered achieve a maximum reward comparable to classical solutions while only requiring a third of the number of trainable parameters and also converging faster. In constructing the hybrid quantum-classical RL model, the internal policy of the agent was replaced by a parametrised quantum circuit. Different architectures were considered for the parametrised quantum circuit. We found that three of the hybrid quantum-classical variants solve the environment. We examined if this performance could be explained by the quantum-specific metrics like expressibility, entanglement capability, or the effective dimension. The latter is particularly interesting since it allows the comparison of classical and quantum-classical architectures. We find however that the performance is not directly linked to these metrics, which could be because some additional factors influencing the performance were missed by this study. Therefore, presently the question on which parametrised quantum circuit to use inside a RL model can only be answered empirically.

Future work has to explore multiple directions. First, this work only considered the simple basis encoding, but it will be interesting to investigate the impact of different encoding techniques on the performance metrics. Furthermore, the PPO algorithm could be replaced by a simpler policy gradient one, where the contribution of the quantum kernel to the solution could possibly be better observed. Additionally, so far, the work was done using only simulations, without any attempts yet to run the algorithms on quantum hardware. This was due to long waiting times and a high number of iterations required between quantum hardware and classical systems. An orthogonal research direction is to consider more sophisticated environments towards more realistic use cases and situations and to therefore bring QRL techniques closer to application.